\documentclass[11pt]{article}
\usepackage{amsmath}
\usepackage{graphicx,psfrag,epsf}
\usepackage{enumerate}
\usepackage{natbib}
\usepackage{amssymb}
\usepackage{amsthm}
\usepackage{fullpage}
\usepackage{url} 

\newcommand{\blind}{0}

\newtheorem{la}{Lemma}
\newtheorem{theorem}{Theorem}

\def\n{\nonumber}

\def\beq{\begin{equation}}
	\def\eeq{\end{equation}}
\def\beqr{\begin{eqnarray}}
	\def\eeqr{\end{eqnarray}}
\def\beqrs{\begin{eqnarray*}}
	\def\eeqrs{\end{eqnarray*}}
\def\bet{\begin{theorem}}
	\def\eet{\end{theorem}}
\def\bel{\begin{lemma}}
	\def\eel{\end{lemma}}
\def\bep{\begin{proposition}}
	\def\eep{\end{proposition}}
\def\bg{\begin{figure}[tbph]\begin{center}}
		\def\eg{\end{center}\end{figure}}

\def\bc{\begin{center}}
	\def\ec{\end{center}}

\def\I{{\bf I}}

\def\mR{\mathbb{R}}

\def\mM{\mathcal M}

\def\be{\begin{equation}}
	\def\ee{\end{equation}}
\def\ben{\begin{equation*}}
	\def\een{\end{equation*}}
\def\bea{\begin{eqnarray}}
	\def\eea{\end{eqnarray}}

\def\bda{\begin{eqnarray*}}
	\def\eda{\end{eqnarray*}}

\newcommand{\bm}{\boldsymbol}

\def\be{\begin{equation}}
	\def\ee{\end{equation}}
\def\ben{\begin{equation*}}
	\def\een{\end{equation*}}
\def\bea{\begin{eqnarray}}
	\def\eea{\end{eqnarray}}

\def\bda{\begin{eqnarray*}}
	\def\eda{\end{eqnarray*}}

\allowdisplaybreaks[4]

\usepackage{lastpage}

\begin{document}

\title{Maximum-of-Differences Test for Comparing Multivariate K-Sample Distributions}

\if0\blind
{
  \title{\bf Maximum-of-Differences Test for Comparing Multivariate K-Sample Distributions}
  \author{
   Wei Lan  \\
	School of Statistics and Center of Statistical Research\\
	Southwestern University of Finance and Economics\\
and\\
 Long Feng \\
School of Statistics and Data Science, Nankai University\\
and\\
	Runze Li \\
 Department of Statistics, Pennsylvania State University\\
 and\\
 Chih-Ling Tsai \\
Graduate School of Management, University of California, Davis\\
}
  \maketitle
\fi

\if1\blind
{
  \bigskip
  \bigskip
  \bigskip
  \begin{center}
    {\LARGE\bf Maximum-of-Differences Test for Comparing Multivariate K-Sample Distributions}
\end{center}
  \medskip
} \fi

\begin{abstract}
Comparing $K$-sample distributions is a fundamental problem in
data science that arises in a wide variety of
fields and applications. In this article,
we introduce a maximum-of-differences approach to make such comparisons.
Specifically, we
first calculate  the pairwise
distances from the pooled observations of the $K$ samples. We then define
the two observations as connected if their distance is less than a pre-specified threshold
value. For
each observation, we  next calculate the ``within" and
the ``between" probabilities  associated with these two types of
connections for the given observation, i.e., with other observations
within the same sample and between the given observation and the observations
in other samples. Subsequently, we propose a
maximum-of-differences  (MOD) test that finds the maximum value
among the standardized squared differences between the ``within"
and  the ``between" probabilities of all observations. Accordingly, the
proposed test is not only applicable to multivariate data with $K$
samples, but can also be extended to multivariate regression models.
Furthermore, we obtain the covariance-adjusted (CA) version of the
MOD (CA-MOD) test, which converges to the Type I extreme value
distribution under some conditions.
Moreover, we demonstrate the asymptotic properties of the two tests under both the null and alternative hypotheses.
The  performance and usefulness of the tests are illustrated via simulation studies and real examples.
\end{abstract}

\noindent%
\vfill

\section{Introduction}

The comparison of $K$ sample distributions is a fundamental problem in  data science that arises in a wide variety of applications across various fields
	such as bioinformatics, economics, finance and machine learning, etc.
	For example, in the area of transfer learning, it is essential to assess the closeness of source data resembles the target data in distributions;
	in database attribute matching, it is desirable to merge databases containing multiple fields with similar distributions of entries;
	in economics, it is critical to test the distributional treatment effects for policy evaluation;
	in the context of gene set analysis (GSA), it is important to test the distribution of gene expression data for $K$ different disease groups.
Due to its practical usage, a stream of research has been devoted in
this direction by generalizing the problem from a fixed dimension to a divergent dimension
(see, e.g., Bai and Saranadasa, 1996;  Zhong et al., 2013), from Euclidean data to non-Euclidean data, such as discrete, network and functional data, (see, e.g., Eagle et al., 2009; Chen and Zhang, 2013; Chen et al., 2018; Wynne and Duncan, 2022),
and from the two-sample comparison problem to the comparison of $K>2$
samples (see, e.g., Mukhopadhyay and Wang, 2020; Mukherjee et al., 2022; Zhang et al., 2024).

In practice, there are three possible approaches that can be used for $K$-sample comparison.
The first is the Hotelling type test statistic and its extensions, which accommodate high-dimensional problems (see, e.g., Bai and Saranadasa, 1996; Li and Chen, 2012; Zhong et al., 2013; Cai et al., 2014; Xue and Yao, 2020).
These tests rely on the assumption that the distribution of observations in the $K$ samples
is parameterized by its mean and a covariance matrix, such as a multivariate normal distribution. In this setting,
the $K$-sample comparison problem is equivalent to testing the equality of the $K$ sample means (or covariances). Hence, Hotelling type tests
are not applicable for the comparison of distributions.
The second method is the classical non-parametric approach, such as the Kolmogorov-Smirnov test, the Wilcoxon test and Wald-Wolfowitz's runs test (Bickel, 1969; Oja and Randles, 2004); see details in the survey article of Gibbons and Chakraborti (2011). These tests are useful and insensitive to distribution assumptions, but they may not perform  well with multivariate data (Jiang et al., 2015; Heller et al., 2016).

The third approach for comparing $K$ samples is the graph-based non-parametric method.
For $K=2$,
Friedman and Rafsky (1979) generalized
Wald-Wolfowitz's runs test to multivariate data.
This method involves creating a minimum spanning tree (MST) based on the pairwise distances among the pooled observations, and then counting
the number of edges that connect nodes (observations) of different samples.
Recently, Chen and Friedman (2017) proposed a modified test
to improve powers either for location or for scale alternatives.
Subsequently, Chen et al. (2018)
introduced  a weighted version of the test to take into account unequal sample sizes.
For $K>2$, Mukhopadhyay and Wang (2020) and Mukherjee et al. (2022) more
recently employed
the graph-based non-parametric approach to propose distribution-free multisample test based on optimal matchings and to develop the GLP test  based on the LP graph kernel, respectively. With an appropriately defined distance measure,
the above tests can be applied to some non-Euclidean data such as discrete and network data with high dimensions. It is of interest to note that
this type of test can be more powerful than the classical non-parametric methods.

In real applications, multivariate data can be associated to some
explanatory variables; see, e.g., the market factor in the notable capital asset pricing model (Sharpe, 1964) and an empirical example in Section 4.
To test the distribution functions with covariates,
Andrews (1997) introduced a conditional Kolmogorov test for distribution functions with covariates.
In addition, Li et al. (2009) proposed a kernel-based test for equality
of distributions to accommodate mixed continuous/discrete variables, and
then extended their result to
the case for testing the equality of two distributions with covariates.
However, the above two tests
are only applicable for fixed dimensional data with $K=2$. Recently, Zhang et al. (2024) constructed a test based on a maximum mean discrepancy for testing
equal distributions of several high-dimensional samples in separable metric spaces. However, their test is not designed for multivariate regression models with $K$ samples.

Since the articles mentioned in the above three approaches mainly focus on multivariate data with $K$-sample comparisons,
this motivates us to
develop a novel approach
that can not only be applied to multivariate data with $K$ samples in both fixed and high dimensions, but can also be directly extended to multivariate regression models with $K$ samples.
To this end, we propose the maximum-of-differences (MOD) test, which is constructed via the following three steps.
\begin{itemize}
	\item [(I.)] Calculate the pairwise distances between
	the pooled observations from the $K$ samples. Then, for
	any two observations, they are defined to be ``connected" if their distance is less than a pre-specified threshold value.
	
	\item [(II.)] For each observation, calculate
	the ``within" and the ``between" probabilities  associated with these two types of connections, i.e., the connections with other observations within
	the same sample  and the connections with the observations in other samples.
	Then compute the difference between the ``within" and the  ``between" probabilities for each observation.
	
	\item [(III.)] Under the null hypothesis that the $K$ samples have equal
	distributions of observations,
	the probability of any two observations being connected is the same. Accordingly, we expect that the difference between
	the ``within" and the ``between" probabilities evaluated in step (II) is close to 0 for each observation.
	This motivates us to
	propose the maximum-of-differences (MOD)
	test, which finds the maximum value among the standardized squared differences between the ``within" and  the ``between" probabilities of all observations.\end{itemize}
	
	We show that, under the null hypothesis,  the MOD test converges to
	the maximum of the $K$ correlated weighted sum of the standard normal and the Type I
	extreme value distributions. Although the
	asymptotic distribution of MOD does not have an analytical form, its critical
	values can be obtained effectively using a numerical method.
	To alleviate the computational inconvenience, however, we propose a modified version of the MOD test.
	Specifically, we profile the covariance structure of the differences before maximization. We name it the covariance-adjusted MOD (CA-MOD) test.
	After the covariance profiling procedure, the elements to be maximized
	are expected to be weakly correlated. We then demonstrate that CA-MOD converges to the Type I extreme value distribution, which can be  used to find the critical value of the CA-MOD test.
	We demonstrate that the proposed two tests are consistent
	under the alternative hypothesis.
	Note that there is no constraint imposed  on the dimension of multivariate data and the number of samples in constructing MOD and CA-MOD.
	Hence, our proposed tests are applicable to multivariate data with
$K\geq 2$ in both fixed and high dimensions.
	We further demonstrate that our method can be directly applied for
	multivariate regression with some restrictions on the size and dimensionality of the data.


	To illustrate the finite sample performance of  the proposed tests, we conduct simulation studies.
	We find that both  MOD and CA-MOD tests control the size well.
	In addition, CA-MOD
	is more powerful than MOD for the covariance and distribution shift alternatives, while the two tests are mostly comparable for
	mean shift alternatives.
	For $K=2$, both tests are comparable with commonly used graph-based methods in testing means when $p$ is small, and they  are both inferior to
	the tests of Chen and Friedman (2017) and Chen et al. (2018) when $p$ is large. However, they are mostly superior to those tests in testing covariances and distributions.
	Furthermore, MOD and CA-MOD perform well  for $K=6$. In general,
	they are superior to the approach of
	Mukherjee et al. (2022) in testing covariances and distributions,
	while they are inferior to (or comparable to) the method of Mukherjee et al. (2022) in testing means.
	Finally, we demonstrate that MOD and CA-MOD  perform well in multivariate regression with $K=2$ and $K=6$.

	The remainder of this article is organized as follows. Section 2 introduces the MOD
	and CA-MOD tests and obtains
	their asymptotic distributions under the null and the alternative hypotheses, respectively.
	Section 3 extends both tests to the multivariate regression setting. Section 4 presents  simulation
	studies and a real data analysis.
	Finally,  Section 5 concludes the article with a short discussion. All theoretical proofs are relegated to the supplementary material.

	\section{Methodology and theoretical results}

	\subsection{Hypothesis Test}

	Let $X^{(k)}_i\in\mR^p$ ($i=1, \cdots, n_k$) be the independent and identically (iid) observations from $F_k$ for $k=1, \cdots, K$.
	The aim of this paper is to test whether the $F_k$s are equal for $k=1, \cdots, K$. Hence, we consider the following hypotheses:
	\beq H_0: F_1=\cdots=F_K \mbox{~v.s.~} H_1:  F_k\not=F_l \mbox{~for some~} k\not=l. \label{hypotheis_unreg}
	\eeq

	To construct the test statistic,
	in the remainder of this paper, with a slight abuse of notation,
	we denote
	$X=(X_1^{(1)}, \cdots, X_{n_1}^{(1)}, \cdots, X_1^{(K)}, \cdots, X_{n_K}^{(K)})=(X_1,\cdots,X_n)\in\mR^{p\times n}$ as the matrix of
	pooled observations, where
	$X_i\in\mR^p$  is the $i$-th column of $X$, for $i=1, \cdots, n$ and $n=n_1+\cdots+n_K$.
	By definition,
	for any $k=1,\cdots, K$ and $l=n_0+\cdots+n_{k-1}+1, \cdots, n_0+n_1+\cdots+n_k$ with $n_0=0$, $X_l$ comes from the $k$-th sample.
	We define $\mathcal{C}_k=\{n_0+\cdots+n_{k-1}+1, \cdots, n_0+\cdots+n_k\}$
	as the collection of indices associated with the observations in the $k$-th sample.
	In addition, for each $i=1,\cdots, n$, we define $g_i$ as the sample that the observation $i$ belongs to. Thus,
	$l\in\mathcal{C}_{g_l}$ means that
	$g_l=k$, for some $k=1,\cdots, K$, and $l=n_0+\cdots+n_{k-1}+1, \cdots, n_0+\cdots+n_k$.

	Under the null hypothesis in (\ref{hypotheis_unreg}),
	the $K$ samples have equal distribution. Accordingly, the $X_i-X_j$ have equal distribution for any $i,j=1,\cdots, n$, $i\not=j$. This implies that
	$\|X_i-X_j\|$ have the same distribution for any $i,j=1,\cdots, n$ and $i\not=j$, where $\|\cdot\|$ is the $L_2$ norm.
	To characterize the distribution of $\|X_i-X_j\|$,
	for any given pre-specified threshold value $\tau$, we define the two observations $i$ and $j$ are ``connected" if $\|X_i-X_j\|\leq\tau$.
	Then, under the null hypothesis of $H_0$, any two ``connected" observations $i$ and $j$ ($i\not=j$) in $K$ samples
	have equal probability $P(\|X_{i}-X_{j}\|\leq\tau)$. In other words, under the null hypothesis,  the same $K$ distributions
	yield the same probability $P(\|X_i-X_j\|\leq\tau)$ for the given $\tau$,
	In addition,
	under the alternative hypothesis, the observations ``within'' and ``between" samples  are connected with likely
	unequal probabilities.
	The different structural information under the null and alternative hypotheses motivates us to
	compare the
	differences between the ``within" and the ``between" probabilities of all observations to form the test statistic.
	
	\subsection{Maximum-of-Differences Test}
	
	For any observation $i=1,\cdots, n$, we calculate the  ``between" and the ``within" probabilities  as
	$\hat{p}_i^{bet}=\frac{1}{n-n_{g_i}}\sum_{j\not\in\mathcal{C}_{g_i}}I(||X_i-X_j||\leq\tau)$ and
	$\hat{p}_i^{in}=\frac{1}{n_{g_i}-1}\sum_{j\in\mathcal{C}_{g_i}, j\not=i}I(||X_i-X_j||\leq\tau)$, respectively,
	where  $I(\cdot)$ is an indicator function and $n_{g_i}$ is the size of $\mathcal{C}_{g_i}$.
	Under the null
	hypothesis of (1),
	we have $E(\hat{p}_i^{bet}-\hat{p}_i^{in})=0$ and $\mbox{var}(\hat{p}_i^{bet}-\hat{p}_i^{in})=(\frac{1}{n-n_{g_i}}+\frac{1}{n_{g_i}-1})\{p_0(1-p_0)-p_{12}\}$,
	where $p_0=P(\|X_{i}-X_{j}\|\leq\tau)$, $p_{12}=E(\delta_{im}\delta_{ij})$ for any $i\not=j\not=m$
	and $\delta_{im}=I(\|X_i-X_m\|\leq\tau)-p_0$.
	We then maximize the standardized of the squared difference between the ''within" and the  ``between" probability across all observations and  obtain
	the following MOD  test statistic,
	\beq \label{test_unreg}
	T_{\tau}=\max_{1\le i\le n} T_{i,\tau}^2=\max_{1\le i\le n} \frac{(\hat{p}_i^{bet}-\hat{p}^{in}_i)^2}{(\frac{1}{n-n_{g_i}}+\frac{1}{n_{g_i}-1})\big\{\hat p_0^{(i)}(1-\hat p_0^{(i)})-\hat{p}^{(i)}_{12}\big\}}, \mbox{~where~}
	\eeq
	\[T_{i,\tau}=\frac{\hat{p}_i^{bet}-\hat{p}^{in}_i}{[(\frac{1}{n-n_{g_i}}+\frac{1}{n_{g_i}-1})\{\hat p_0^{(i)}(1-\hat p_0^{(i)})-\hat{p}^{(i)}_{12}\}]^{1/2}}.\]
	Under the null hypothesis of $H_0$, $\hat p_0^{(i)}=(n-1)^{-1}\sum_{j\not=i}I(\|X_i-X_j\|\leq\tau)$ is the consistent estimator of
	$p_0$, $\hat{p}^{(i)}_{12}=\{n(n-1)\}^{-1}\sum_{m=1}^n\sum_{j\not=m\not=i}\hat\delta^{(i)}_{im}\hat\delta^{(i)}_{ij}$ is the consistent estimator of $p_{12}$,
	and $\hat\delta^{(i)}_{im}=I(\|X_i-X_m\|\leq\tau)-\hat p^{(i)}_0$.
	We expect that
	$T_{\tau}$ is small under the null hypothesis, whereas it is large under the alternative hypotheses.
	Note that we construct the statistic,  $T_{i,\tau}$, by substituting the unknown parameters, $p_0$ and $p_{12}$,
	with their corresponding estimators, $\hat p_0^{(i)}$ and $\hat p_{12}^{(i)}$, for each  $i=1,\cdots, n$.
	This approach is not only computationally  convenient, but also
	mitigates some possible correlations among the test statistics $T_{i,\tau}$s.
	The resulting test statistic, MOD, performs reasonably well based on simulation studies in Section 4.

	To derive the asymptotic distribution of $T_{\tau}$, we need to find the covariance structure of the $T_{i,\tau}$s for $i=1, \cdots, n$.
	To this end, define
	\[Q_{i,\tau}=\frac{\hat{p}_i^{bet}-\hat{p}^{in}_i}{\{(\frac{1}{n-n_{g_i}}+\frac{1}{n_{g_i}-1})(p_0(1-p_0)-p_{12})\}^{1/2}},\] and
	let $\Sigma=(\sigma_{ij})\in\mR^{n\times n}$ be the covariance matrix of $(Q_{1,\tau}, \cdots, Q_{n,\tau})^\top$.
	In addition, let $p_{22}=p_0(1-p_0)$ for ease of  notation.
	Then, under the null hypothesis of $H_0$,
	algebraic calculation implies that
	\[\left\{
	\begin{aligned}
		\sigma_{ij}&=\frac{\{\frac{1}{n-n_{g_{i}}}+\frac{1}{n_{g_{i}-1}}\}p_{12}-\frac{3p_{12}-p_{22}}{(n_{g_{i}}-1)^2}}{(\frac{1}{n-n_{g_{i}}}+\frac{1}{n_{g_{i}}-1})(p_{22}-p_{21})}, \mbox{~if $i$ and $j$ belong to the same sample;~}\\
		\sigma_{ij}&=\frac{(n_{g_{i}}-1)^{1/2}(n_{g_{j}}-1)^{1/2}\{(p_{22}-(n+2)p_{12})\}}{(n-n_{g_{i}})^{1/2}(n-n_{g_{j}})^{1/2}(n-1)(p_{22}-p_{12})}, \mbox{~otherwise.~}\\
	\end{aligned}
	\right.
	\]
	%
	%
	%
	The detailed derivation of $\sigma_{ij}$ is given in
	Section S.3 of the supplementary material.
	Moreover, let $\hat\Sigma$ be the estimator of $\Sigma$ by replacing $p_0$, $p_{12}$, $p_{22}$ and $\delta_{im}$ with
	$\hat p_0=\{n(n-1)\}^{-1}\sum_{i=1}^n\sum_{m\not=i}I(\|X_i-X_m\|\leq\tau)$,
	$\hat p_{12}=\{n(n-1)(n-2)\}^{-1}\sum_{i=1}^n\sum_{m\not=i}\sum_{j\not=m\not=i}\hat\delta_{im}\hat\delta_{ij}$,
	$\hat p_{22}=\hat p_0(1-\hat p_0)$
	and $\hat\delta_{im}=I(\|X_i-X_m\|\leq\tau)-\hat p_0$, respectively.
	Since $\hat\Sigma$ has a group-wise structure of $K$ groups, it can be calculated effectively. Furthermore, we can demonstrate that $\hat\Sigma$ is the consistent estimator of $\Sigma$.
	
	To establish the theoretical properties of $T_{\tau}$, we introduce the following condition:
	\begin{itemize}
		\item[(C1)] For any $k=1,\cdots, K$ and $K<\infty$, assume that $n_k/n\rightarrow \gamma_k$ for some positive constant $\gamma_k$. In addition, assume that
		there exist finite positive constants $\gamma_{\min}$ and $\gamma_{\max}$ such that $0<\gamma_{\min}<\min_k \gamma_k\leq \max_k \gamma_k<\gamma_{\max}<1$ for any $n$.
	\end{itemize}
	\noindent
	Condition (C1) assumes that the size of each sample is of the same order even asymptotically;
	thus, there are
	sufficient observations in each sample.
	Based on this condition,
	we obtain the asymptotic distribution of $T_{\tau}$ given below.
	
	\bet
	Assume Condition (C1) is satisfied. Under $H_0$, for any $0<p_0<1$
	and $x\in\mR^{+}$, we have that
	\begin{align} \label{t11}
		P\big\{T_{\tau}<x\big\}-P(\max_{1\le i\leq n} Z_i^2<x)\rightarrow 0,
	\end{align}
	as $n\to \infty$, where $Z=(Z_1, \cdots, Z_n)^\top$ is a multivariate normally distributed random variable with mean zero and covariance matrix $\Sigma$.
	\eet
	\noindent In Theorem 1, we require $0<p_0<1$. Recall that $p_0=P(\|X_i-X_j\|\leq\tau)$, $i,j=1,\cdots,n$, $i\ne j$, and the dimensions of $X_i$ and $X_j$ are $p$.
	Thus, $\tau$  implicitly depends on the sample size $n$ and the dimension of $X_i$, $p$; it is not ``fixed".
	To make the above theorem practically useful, we can replace $\Sigma$  by its consistent estimator $\hat\Sigma$.
	In  Theorem 1, we can express $\max_{1\le i\leq n} Z_i^2=\max_{1\leq k\leq K} \max_{i\in \mathcal{C}_k} Z_i^2$.  Note that, under the null hypothesis,
	for all $i$ belonging to the sample $\mathcal{C}_k$, the $Z_i$s have the same covariance \[\bar\rho^{(k)}=\frac{\{\frac{1}{n-n_{g_{i}}}+\frac{1}{n_{g_{i}-1}}\}p_{12}-\frac{3p_{12}-p_{22}}{(n_{g_{i}}-1)^2}}{(\frac{1}{n-n_{g_{i}}}+\frac{1}{n_{g_{i}}-1})(p_{22}-p_{21})},\] where
	$n_{g_i}$ is the size of the sample $\mathcal{C}_{g_i}=\mathcal{C}_k$.
	Then, by the result of Lemma 1 in Section S.1 of the supplementary material,
	$\max_{i\in \mathcal{C}_k} Z_i^2$ converges to a mixture distribution of the standard  normal and extreme value distributions $\sqrt{\bar\rho^{(k)}} N(0,1)+\sqrt{1-\bar\rho^{(k)}} ED$,
	where $N(0,1)$ is the standard normal distribution and $ED$ is the Type I extreme value distribution with
	cumulative distribution function
	$\exp\{-\pi^{-1/2}\exp(-x/2)\}$.
	Accordingly,
	$\max_{1\le i\leq n} Z_i^2$ converges to the maximum of the $K$ correlated weighted sums of the standard normal and extreme value distributions.
	The critical values can be obtained
	via the following replication method with two steps.
		\begin{itemize}
			\item [Step (I.)] At the $b$-th step of the replication procedure of $b=1,\cdots, B$, we independently generate $N$ realizations of the $n$ iid observations from the multivariate normal distribution with mean zero and covariance matrix $\hat\Sigma$, and denote them as $O_b^{(l)}=(O_{b1}^{(l)}, \cdots, O_{bn}^{(l)})^\top$ for
			$l=1,\cdots, N$.
			\item [Step (II.)] We calculate $J_b^{(l)}=\max_i O_{bi}^{(l)2}$, and let $J_{b,\alpha}$ be the empirical $(1-\alpha)$-quantile
			of $J_b^{(1)}, \cdots, J_b^{(N)}$. Repeat this procedure $B$ times for $b=1,\cdots, B$, then  evaluate the empirical $p$-value of the test statistic,
			$p_{MOD}=B^{-1}\sum_{b=1}^B I(T_{\tau}>J_{b,\alpha})$.
		\end{itemize}
		The validity of this replication method is guaranteed by the law of large numbers (see Lehmann 2004).

	{We next evaluate the power of the test.
		Let $p_{kl,\tau}=P(||X_i-X_j||\leq\tau)$ for any $i\in \mathcal{C}_k$ and $j\in \mathcal{C}_l$,
		and $p_{kls,\tau}=E\big[(I(||X_i-X_j||\leq\tau)-p_{kl,\tau})(I(||X_i-X_t||\leq\tau)-p_{ks,\tau})\big]$ for any  $i\in \mathcal{C}_k, j\in \mathcal{C}_l$, and $t\in\mathcal{C}_s$.
		Under the null hypothesis,  $p_{kl,\tau}=p_0$ and $p_{kls,\tau}=p_{12}$  for any $k, l, s=1,\cdots, K$.
		In addition, to characterize the differences between the ``within" and the ``between" probabilities for each observation,
		let
		\begin{align} \label{power_nu}
			\nu_i=\frac{\big(\sum_{k\not=g_i}\frac{\gamma_{k}}{1-\gamma_{g_i}}p_{g_ik,\tau}
				-p_{g_ig_i,\tau}\big)^2}{\big(\sum_{k=1}^K \gamma_k p_{g_ik,\tau}\big)\big(1-\sum_{k=1}^K \gamma_k p_{g_ik,\tau}\big)-\tilde{\delta}_{12i}}
		\end{align}
		and $\tilde{\delta}_{12i}\doteq\sum_{k,l} \gamma_k\gamma_l\big\{p_{g_ikl,\tau}+(p_{g_ik,\tau}-\sum_{s=1}^K\gamma_sp_{g_is,\tau})(p_{g_il,\tau}-\sum_{s=1}^K\gamma_sp_{g_is,\tau})\big\}$,
		where $\gamma_k$ was defined in Condition (C1). Under the null and alternative hypotheses, we have $\nu_i=0$ and  $\nu_i>0$, respectively. Thus,
		$\nu_i$ can measure the discrepancy between the null and alternative hypotheses.
		We then show the consistency of  the MOD test given below.
		
		\begin{theorem}\label{th2}
			Assume Condition (C1) holds and $0<p_0<1$. Under the alternative hypotheses of $H_1$,
			if $n\max_{1\le i\le n}\nu_i>(4+\kappa) \log n$ for some finite constant $\kappa>0$,
			we have that $P(T_{\tau}>z_{\alpha})\to 1$,  as $n\to \infty$, where $z_{\alpha}$ is the upper $\alpha$-th quantile of $\max_{1\leq i\leq n} Z_i^2$.
		\end{theorem}
		
		By Theorem 2, the power of MOD is closely related to $\max_{1\le i\le n}\nu_i$, which measures the maximum difference between the ``within" and the ``between" probabilities across observations. Accordingly, $\max_{1\le i\le n}\nu_i$ gauges the discrepancy between the null and alternative hypotheses, and MOD is consistent under the alternatives that satisfy
			$n\max_{1\le i\le n}\nu_i>(4+\kappa) \log n$.

\subsection{Tuning Parameter Selection}
		
By Theorem 2, the power of MOD is associated to
$\max_{1\le i\le n}\nu_i$, which  motivates us to
		select the tuning parameter $\tau$ by maximizing $\max_{1\le i\le n} \nu_i$.
		For any $k,l=1,\cdots, K$, let $F_{kl}$ be the cumulative distribution function of $||X_i-X_j||^2$ for $i\in\mathcal{C}_k$, $j\in\mathcal{C}_l$ and $i\not=j$.
		Then, for the given $\tau$, we have $F_{kl}(\tau^2)=P(\|X_i-X_j\|\leq \tau)$. Note that
		$i\in\mathcal{C}_k$ and $j\in\mathcal{C}_l$ indicate $g_i=k$ and $g_j=l$.
		As a result, $F_{g_ig_j}(\tau^2)=P(\|X_i-X_j\|\leq \tau)$ for any $i\not=j$, and $F_{g_ig_i}(\tau^2)=P(\|X_{i}-X_{i_1}\|\leq \tau)$
		for $i, i_1 \in\mathcal{C}_{g_i}$ and $i\not=i_1$.
		In addition, define the inverse function and the density function of $F_{kl}$ as $F^{-1}_{kl}$  and $f_{kl}=F'_{kl}$, respectively.
		It is worth noting that $F_{kl}(\tau^2)=p_{kl,\tau}$, where $p_{kl,\tau}$ was defined above (\ref{power_nu}).
		The reason for introducing  $F_{kl}$
		is that the $F^{-1}_{kl}$ and $f_{kl}$ will be used hereafter.
		
		Under the null hypothesis, $F_{kk}=F_{kl}$ for any $k,l$. In contrast,
		under the alternative hypothesis, $F_{kk}\not=F_{kl}$ for some
		$k\not=l$, which leads to $F_{kk}^{-1}F_{kl}(\tau^2)-\tau^2\not=0$.
		This motivates us to consider a specific alternative hypothesis:
		\begin{align} \label{h11}
			F_{kk}^{-1}F_{kl}(\tau^2)-\tau^2=\omega_{kl} \sqrt{\log n/n},~\text{for all~} \tau ~\text{and}~ k\not=l,
		\end{align}
		where $\omega_{kl}$ are finite constants that do not depend on $\tau$ asymptotically.
		
		We next evaluate $\nu_i$ in (\ref{power_nu}) under the alternative hypothesis (\ref{h11}).
		By the Taylor series expansion, we obtain that
		\begin{align}
			p_{g_ig_j,\tau}-p_{g_ig_i,\tau}=&F_{g_ig_j}(\tau^2)-F_{g_ig_i}(\tau^2)=F_{g_ig_i}(F_{g_ig_i}^{-1}F_{g_ig_j}(\tau^2))-F_{g_ig_i}(\tau^2)\n \\
			=&f_{g_ig_i}(\tau^2)(F_{g_ig_i}^{-1}F_{g_ig_j}(\tau^2)-\tau^2)+O(\log n/n)\n \\
			=& \omega_{g_ig_j} \sqrt{\log n/n}f_{g_ig_i}(\tau^2)+O(\log n/n). \label{alternative_t1}
		\end{align}
		Thus, $p_{g_ig_j,\tau}-p_{g_ig_i,\tau}=O(\sqrt{\log n/n})=o(1)$.
		Applying this result, the two terms in the denominator of $\nu_i$ are such that
		\begin{align}
			&\sum_{k=1}^K \gamma_k p_{g_ik,\tau}\to p_{g_ig_i,\tau} \mbox{~and~} \tilde{\delta}_{12i}\to \sum_{k,l}\gamma_k\gamma_l p_{g_ikl,\tau}.\label{alternative_t2}
		\end{align}
		In addition,  the term
		in the numerator of $\nu_i$ is such that
		\begin{align}
			&\sum_{k\not=g_i}\frac{\gamma_{k}}{1-\gamma_{g_i}}p_{g_ik,\tau}
			-p_{g_ig_i,\tau}=
			f_{g_ig_i}(\tau^2) \sum_{k\not=g_i}\frac{\gamma_{k}}{1-\gamma_{g_i}} \omega_{g_ig_j}+O(\log n/n), \label{alternative_t3}
		\end{align}
		where $ \sum_{k\not=g_i} \frac{\gamma_{k}}{1-\gamma_{g_i}} \omega_{g_ig_j}$ is free of $\tau^2$ asymptotically.

		Under the alternative hypothesis (\ref{h11}), the results of (\ref{alternative_t1})--(\ref{alternative_t3}) indicate that
		finding the maximum of $\max_{1\le i\le n}\nu_i$ asymptotically
		is sufficient to maximize the following function,
		\begin{align*}
			J_i(\tau)=\frac{f^2_{g_ig_i}(\tau^2)}{p_{g_ig_i,\tau}(1-p_{g_ig_i,\tau})-\sum_{k,l}\gamma_k\gamma_l p_{g_ikl,\tau}}
		\end{align*}
		uniformly for any $i$.
		To this end, define $\xi\doteq p_{g_ig_i,\tau}=F_{g_ig_i}(\tau^2)\in (0,1)$,  and we then have $\tau^2={F}^{-1}_{g_ig_i}(\xi)$. Since $F_{g_ig_i}$ is a non-decreasing function of $\tau^2$,
		maximizing the function $J_i(\tau)$ of $\tau^2$ is equivalent to maximize the following function
		\begin{align}\label{jxi}
			\mathcal{J}_i(\xi)=&\frac{f^2_{g_ig_i}(F^{-1}_{g_ig_i}(\xi))}{\xi(1-\xi)-\sum_{k,l}\gamma_k\gamma_l p_{g_ikl,F^{-1/2}_{g_ig_i}(\xi)}}
		\end{align}
		uniformly for any $i$, where  $\xi\in (0,1)$.
		Let $\hat\xi=\arg\max_{\xi \in (0,1)}\mathcal{J}_i(\xi)$ for any $i$.
		Using the above result,  we then choose $\tau^2={F}^{-1}_{g_ig_i}(\hat{\xi})$.
		In practice,  ${F}$ is unknown. Therefore,  we choose $\tau^2=\hat{F}^{-1}_{g_ig_i}(\hat{\xi})$, where $\hat{F}_{g_ig_i}$ is the empirical distribution function of $F_{g_ig_i}$.
		
		To obtain the explicit formula of $\hat\xi$ and characterize the magnitude of $\max_{1\le i\le n}\nu_i$ required in Theorem 2, we consider the following two examples.
		
		Example I: two normal samples with mean shift. We consider
		$X_{1i}\sim N(0,\I_p)$ and  $X_{2j}\sim N(\mu, \I_p)$, where the notation $\sim$ means ``is distributed as." Let $X_{1i}$ and $X_{2j}$ be the observations from samples 1 and 2, respectively,  for $i=1,\cdots, n_1$ and $j=1,\cdots, n_2$. We then  have $||X_{1i_1}-X_{1i_2}||^2\sim N(2p,8p)$ for $i_1, i_2=1,\cdots, n_1$ and $i_1\not=i_2$,
		and $||X_{1i}-X_{2j}||^2\sim N((\mu^2+2)p,8p)$. Employing the Taylor series
		expansion,
		we further obtain that
		\begin{align*}
			F_{11}(\tau^2)-F_{12}(\tau^2)=&\Phi\left(\frac{\tau^2-2p}{\sqrt{8p}}\right)-\Phi\left(\frac{\tau^2-(\mu^2+2)p}{\sqrt{8p}}\right)\\
			=&\phi\left(\frac{\tau^2-2p}{\sqrt{8p}}\right)\frac{\mu^2p}{\sqrt{8p}}\{1+o(1)\},
		\end{align*}
		where $\Phi$ and $\phi$ are the cumulation distribution function and density function of the standard normal distribution, respectively. This, together with (\ref{alternative_t1}), implies that
		$F_{11}^{-1}F_{12}(\tau^2)-\tau^2=\frac{\mu^2p}{\sqrt{8p}}\{1+o(1)\}$.
		As a result, it does not depend on $\tau$ asymptotically and
		satisfies the alternative in (\ref{h11})
		if $\mu^2=O\{\sqrt{\log n/(np)}\}$.
		We can further show that $p_{g_ikl,\tau}=O(p^{-1})$, which tends to 0 as $p$ diverges. Accordingly, the term $\sum_{k,l}\gamma_k\gamma_l p_{g_ikl,F^{-1/2}_{g_ig_i}(\xi)}$ in (\ref{jxi}) is negligible.  This, in conjunction with (\ref{jxi}), implies
		that $\hat{\xi}=0.5$, which  maximizes $\frac{\phi^2(\Phi^{-1}(\xi))}{\xi(1-\xi)}$.
		In addition, under this specific choice $\hat{\xi}=0.5$, one can verify that the theoretical requirement in Theorem 2, $n\max_{1\le i\le n}\nu_i>(4+\kappa) \log n$,
			holds as long as $\mu^4\geq 4\pi (4+\kappa)\log n/(np)$ in this mean shift example.

		Example II: two normal samples with covariance shift. Consider
		$X_{1i}\sim N(0,\I_p)$ and $X_{2j}\sim N(0, (1+\sigma^2)\I_p)$.
		Then, $||X_{1i_1}-X_{1i_2}||^2\sim N(2p,8p)$ and $||X_{1i}-X_{2j}||^2\sim N(2(1+\sigma^2)p,8(1+\sigma^2)^2p)$.
		Employing the Taylor
		expansion, we then obtain that
		\begin{align*}
			F_{11}(\tau^2)-F_{12}(\tau^2)=&\Phi\left(\frac{\tau^2-2p}{\sqrt{8p}}\right)-\Phi\left(\frac{\tau^2-2(1+\sigma^2)p}{\sqrt{8(1+\sigma^2)^2p}}\right)\\
			=&\phi\left(\frac{\tau^2-2p}{\sqrt{8p}}\right)\frac{\tau^2\sigma^2}{\sqrt{8p}(1+\sigma^2)}\{1+o(1)\}\\
			=&\phi\left(\frac{\tau^2-2p}{\sqrt{8p}}\right)\frac{p^{1/2}\sigma^2}{\sqrt{2}(1+\sigma^2)}\{1+o(1)\}.
		\end{align*}
		By (\ref{alternative_t1}), we have $F_{11}^{-1}F_{12}(\tau^2)-\tau^2=\frac{p^{1/2}\sigma^2}{\sqrt{2}(1+\sigma^2)}\{1+o(1)\}$,
		which does not depend on $\tau$ and satisfies (\ref{h11}) if $\sigma^2=O\{\sqrt{\log n/(np)}\}$.
		In addition, we can show that $p_{g_ikl,\tau}=O(p^{-1})$. This, together with (\ref{jxi}), leads to $\hat{\xi}=0.5$, which  maximizes
		$\frac{\phi^2(\Phi^{-1}(\xi))}{\xi(1-\xi)}$ asymptotically. Furthermore,
			under this specific choice $\hat{\xi}=0.5$, one can verify that the theoretical requirement in Theorem 2, $n\max_{1\le i\le n}\nu_i>(4+\kappa) \log n$,
			holds as long as $\sigma^4\geq 4\pi (4+\kappa)\log n/(np)$ in this covariance shift example.

		In the above two examples, we obtain that $\hat{\xi}=0.5$ asymptotically.
		Thus, $\tau$ is set to be the median of the $\|X_i-X_j\|$ for $i,j=1,\cdots, n$, and we recommend this choice in real practice.}
	
	\subsection{Covariance-Adjusted Maximum-of-Differences Test}
	
	Due to the complex covariance structures between the $T_{i,\tau}s$,  the asymptotic distribution of $T_{\tau}$ cannot
	be expressed in an analytical form.
	To alleviate this computational inconvenience, we propose a method of adjusting the covariance structure of the $T_{i,\tau}s$ before the maximization.
	Let the resulting adjustment be $\mM^{adj}_{\tau}=\hat\Sigma^{-1/2}(T_{1,\tau}, \cdots, T_{n,\tau})^\top=(\mM^{adj}_{1,\tau}, \cdots, \mM^{adj}_{n,\tau})^\top$.
	We then obtain the covariance-adjusted maximum-of-differences test statistic, CA-MOD, given below,
	\beq
	T^{adj}_{\tau}=\max_{1\le i\le n} \big(\mM^{adj}_{i,\tau}\big)^2. \label{adjust_test}
	\eeq
	After the covariance adjusting procedure, the elements of $\mM^{adj}_{\tau}$ are  weakly correlated.
	Hence, we expect that $T^{adj}_{\tau}$  converges to a Type I extreme value
	distribution; see the following result.

	\bet
	Assume Condition (C1) is satisfied and $\Sigma$ is positive definite for any $n$.  Under $H_0$, for any $0<p_0<1$ and
	$x\in\mR^{+}$,  we have that
	\[P\Big\{T^{adj}_{\tau}-2\log(n)+\log\log(n)\le x\Big\}\to \exp\Big\{-\frac{1}{\sqrt{\pi}}\exp(-x/2)\Big\}, \]
	as $n\to \infty$.
	\eet
	
	Based on the above theorem, we reject the null hypothesis of $H_0$ at a significant level
	$\alpha$ if $T^{adj}_{\tau}-2\log(n)+\log\log(n)\ge q_{\alpha}$,
	where $q_{\alpha}\equiv-\log(\pi)-2\log\{\log(1-\alpha)^{-1}\}$ is the
	$(1-\alpha)$-quantile of the Type I extreme value distribution with
	the cumulative distribution function
	$\exp\{-\pi^{-1/2}\exp(-x/2)\}$.

	Compared with MOD, CA-MOD requires an additional condition that $\Sigma$ is positive definite for any $n$. To illustrate this condition, we consider a special case: $K=2$,
	$p_{0}=0.5$, $n_1=n_2$, and
	$X_i$ follows a multivariate normal distribution with mean 0 and covariance matrix $I_p$, for $i=1,\cdots, n$.
	After simple calculations, we have that $p_{12}=\frac{1}{\pi p}\{1+o(1)\}$ and $p_{22}=0.25$.
	Then, $\sigma_{ij}=\frac{4}{\pi p-4} \{1+o(1)\}$ if $i$ and $j$ belong to the same group and
	$\sigma_{ij}=-\frac{4}{\pi p-4} \{1+o(1)\}$ otherwise.
	One can further verify that $\lambda_{\max}(\Sigma)$ and $\lambda_{\min}(\Sigma)$ are of orders $1+O(n\sigma_{ij})=1+O(n/p)$ and  $1-O(p^{-1})$, respectively. As $p$ gets large, we always have $\lambda_{\min}(\Sigma)>0$ and thus $\Sigma$ is positive definite for any $n$.

	Define $\tilde{p}_{22}=\tilde p_{0}(1-\tilde{p}_{0})$, where $\tilde{p}_{0}=\sum_{k=1}^K\sum_{l=1}^K \gamma_k\gamma_l p_{kl,\tau}$,
	$\tilde{p}_{12}=\sum_{i=1}^K \gamma_i \tilde{\delta}_{12i}$, and $\tilde{\delta}_{12i}$ was defined above Theorem 2.
	Under the alternative hypothesis, we can prove that $\hat{p}_{0}\to \tilde{p}_{0}$ and $\hat{p}_{12}\to \tilde{p}_{12}$.
	Define $\Sigma^*=(\sigma_{ij}^*)$, where $\sigma_{ij}^*$ are obtained by replacing $p_{12},p_{22}$ in $\sigma_{ij}$ with
	$\tilde{p}_{12}, \tilde{p}_{22}$, respectively, for $i,j=1,\cdots,n$. That is, $\Sigma^*$ is the counterpart of $\Sigma$ under the alternative hypothesis. Then,
	we have $\hat{\Sigma}\to \Sigma^*$ under the alternative hypothesis. In addition,  define $\omega=(\omega_1,\cdots,\omega_n)^\top={\Sigma^*}^{-1/2}(\nu_1^{1/2},\cdots,\nu_n^{1/2})$,
	and the consistent property of CA-MOD is given below.
	
	\begin{theorem}\label{th4}
		Assume Condition (C1) is satisfied and $0<p_0<1$, and $\Sigma^*$ is positive definite for any $n$. Then, if
		$n\max_{1\le i\le n}\omega_i^2>(4+\tilde\kappa)\log n$ for
some finite constant
		$\tilde\kappa>0$, we have that
		$P\left(T^{adj}_{\tau}-2\log(n)+\log\log(n)\ge q_{\alpha}\right)\to 1$.
	\end{theorem}

	\noindent\textbf{Remark 1:}
	It is worth noting that $T_{i,\tau}$ is a summation of the correlated random variables.
	Thus,
	the classical result based on the maximum of independent summations cannot be  applied
	to obtain the limiting distribution of the MOD and CA-MOD tests. To overcome this challenge,
	we employ two expansions known as the
	Slepian interpolation  and Stein's
	leave-one-out (see e.g., Chernozhukov et al. 2013) to first extract the correlated terms, and then show that the difference between the extracted form and the raw form is negligible.
	Based on this result, we are able to
	obtain the asymptotic distribution of the two proposed tests.
	Their detailed proofs are presented in Sections S.2-S.5 of the supplementary material, respectively.

	\noindent\textbf{Remark 2:} It is worth noting that the test statistics $T_{i,\tau}$ and $\mM^{adj}_{i,\tau}$
	involved in (\ref{test_unreg}) and (\ref{adjust_test}), respectively, are both asymptotically standard normal for $i=1, \cdots, n$
	under the null hypothesis $H_0$.
	Hence, the asymptotic distributions of $T_{i,\tau}$ and $\mM^{adj}_{i,\tau}$ are not related to $\tau$.
	Instead, to make MOD and CA-MOD practically useful, one needs to specify the
	hyper-parameter $\tau$ {\it{a priori}}. Note that the test statistic (\ref{test_unreg}) involves
	the term $\hat p^{(i)}_0(1-\hat p^{(i)}_0)-\hat p^{(i)}_{12}$ in the denominator for each $i=1, \cdots, n$.
	
	According to the discussion in Section 2.3,
	we suggest choosing $\tau$ to be the median of the $\|X_i-X_j\|$ across $i,j=1,\cdots, n$, for $i\ne j$.
	This choice not only maximizes the power under a special type of alternative hypothesis, but also
	avoids the $\hat p_0^{(i)}$s being extreme small (i.e., close to 0) or large (i.e., close to 1).
	Thus, numerical calculations of the MOD and CA-MOD tests are steady.
	For the sake of illustration, we conduct simulation studies in Section 4 by choosing three different $\tau$s (i.e., the empirical
	25\%, 50\% and 75\% quantiles of the $\|X_i-X_j\|$ across $i,j=1,\cdots, n$, for $i\ne j$).
	The results indicate that the three different $\tau$s yield comparable results, while
	the median performs the best in most cases.

	\noindent\textbf{Remark 3:}
	The MOD and CA-MOD tests only rely on the binary variables
	$I(||X_i-X_j||\leq\tau)$, for $i,j=1,\cdots, n$ and $i\ne j$, which do not require
	any restrictions on the orders of $p$. In addition, these binary variables are not directly related to the
	distributions of $X_i$ for $i=1,\cdots, n$. Thus, the proposed two tests are applicable for fat-tailed data.
	Moreover, the theoretical results in Theorems 1--2 are not affected by the choice
	of distance measure.
	Simulation results in Section 4  show that MOD and CA-MOD perform satisfactorily based on the $L_2$ distance measure.
	Hence, we leave the selection of
	the optimal distance measure for possible future research; see the concluding remarks.

	In practice,  $K$-sample comparisons are important in linear regression models. Hence, we extend  both the MOD and CA-MOD tests in Section 3. We also compare the finite sample performance of
	these two tests via Monte Carlo studies in Section 4.

	\section{Testing in a regression setting}

	In real applications, multivariate data can be associated to some explanatory variables
	(see, e.g., Andrews, 1996). For example, in finance data, the stock
	returns are usually driven by some systematic factors (i.e., explanatory variables or covariates) such as
	the market factor in the notable capital asset pricing model (Sharpe, 1964); see also an
	empirical example in Section 4. This motivates us to conduct $K$-sample
	comparisons after adjusting for the covariates. To this end, we consider the following
	multivariate regression model
	\beq X_i^{(k)}=\beta^{(k)}W^{(k)}_i+\epsilon_i^{(k)}, \label{reg_model}\eeq
	for $i=1,\cdots, n_k$ and $k=1,\cdots, K$, where $W^{(k)}_i\in\mR^d$ is the $d$-dimensional covariate ($d<\infty$), and $\beta^{(k)}\in\mR^{p\times d}$ is the regression
	coefficient matrix of the $k$-th sample. In addition, $\epsilon_i^{(k)}\in\mR^p$ is the random error that is independent of $W^{(k)}_i$, and it is assumed to be
	independent and identically distributed with  distribution function $F_k^{\epsilon}$.
	Then the null and alternative hypothesis for testing the $K$ sample distributions are, respectively,
	\beq \label{hypotheis_reg}H_{0}: F^{\epsilon}_1=\cdots=F^{\epsilon}_K \mbox{~v.s.~} H_{1}: F^{\epsilon}_k\not=F^{\epsilon}_l \mbox{~for some~} k\not=l.\eeq
	It is worth noting that testing the equal distribution of random errors $\epsilon_i^{(k)}$ can be used to test the error assumption of the
	normality in a regression model.

	To construct the test statistic, we fit model (\ref{reg_model}) with the ordinary least squares method and obtain the
	estimated error $\hat\epsilon_i^{(k)}=\big\{I_{n_k}-W^{(k)}(W^{(k)\top}W^{(k)})^{-1}W^{(k)\top}\big\}X_i^{(k)}$ with
	$W^{(k)}=(W_1^{(k)}, \cdots, W_{n_k}^{(k)})^\top\in\mR^{n_k\times d}$. Subsequently, we define
	$\epsilon=(\epsilon_1^{(1)}, \cdots, \epsilon_{n_1}^{(1)}, \cdots, \epsilon_1^{(K)}, \cdots, \epsilon_{n_K}^{(K)})=(\epsilon_1, \cdots, \epsilon_n)\in\mR^{p\times n}$,
	$\hat\epsilon=(\hat\epsilon_1^{(1)}, \cdots, \hat\epsilon_{n_1}^{(1)}, \cdots, \hat\epsilon_1^{(K)}, \cdots, \hat\epsilon_{n_K}^{(K)})=(\hat\epsilon_1, \cdots, \hat\epsilon_n)\in\mR^{p\times n}$,
	$\hat{p}_i^{\hat\epsilon, bet}=\frac{1}{n-n_{g_i}}\sum_{j\not\in\mathcal{C}_{g_i}}I(||\hat\epsilon_i-\hat\epsilon_j||\leq\tau)$,
	$\hat{p}_i^{\hat\epsilon, in}=\frac{1}{n_{g_i}-1}\sum_{j\in\mathcal{C}_{g_i}}I(||\hat\epsilon_i-\hat\epsilon_j||\leq\tau)$,
	$\delta^{\epsilon}_{ik}=I(\|\epsilon_i-\epsilon_k\|\leq\tau)-p^{\epsilon}_0$, $p^{\epsilon}_{12}=E(\delta^{\epsilon}_{ik}\delta^{\epsilon}_{ij})$ for any $i\not=k\not=j$,
	$p^{\epsilon}_{22}=p^{\epsilon}_0(1-p^{\epsilon}_0)$ and
	$p_0^{\epsilon}=P(\|\epsilon_i-\epsilon_k\|\leq\tau)$.
	Under the null hypothesis of $H_0$, $\hat p_0^{\hat\epsilon,(i)}=(n-1)^{-1}\sum_{k\not=i}I(\|\hat\epsilon_i-\hat\epsilon_k\|\leq\tau)$ is the consistent estimator of
	$p^{\epsilon}_0$, and $\hat{p}^{\hat\epsilon,(i)}_{12}=\{n(n-1)\}^{-1}\sum_{k=1}^n\sum_{j\not=k\not=i}\hat\delta_{ik}^{\epsilon,(i)}\hat\delta_{ij}^{\epsilon,(i)}$ is the consistent estimator of $p^{\epsilon}_{12}$,
	where $\hat\delta^{\hat\epsilon,(i)}_{ik}=I(\|\hat\epsilon_i-\hat\epsilon_k\|\leq\tau)-\hat p_0^{\hat\epsilon,(i)}$.

	Analogous to the procedure for testing (\ref{hypotheis_unreg}),
	we define
	\[T^{\hat\epsilon}_{i,\tau}=\frac{\hat{p}_i^{\hat\epsilon,bet}-\hat{p}^{\hat\epsilon,in}_i}{[(\frac{1}{n-n_{g_i}}+\frac{1}{n_{g_i}-1})\{\hat p^{\hat\epsilon,(i)}_0(1-\hat p^{\hat\epsilon,(i)}_0)-\hat{p}^{\hat\epsilon,(i)}_{12}\}]^{1/2}}, \mbox{~and~}\]
	\[Q^{\epsilon}_{i,\tau}=\frac{\hat{p}_i^{\hat\epsilon,bet}-\hat{p}^{\hat\epsilon,in}_i}{[(\frac{1}{n-n_{g_i}}+\frac{1}{n_{g_i}-1})\{p^{\epsilon}_0(1-p^{\epsilon}_0)-p^{\epsilon}_{12}\}]^{1/2}},\] and let
	$\Sigma^{\epsilon}=(\sigma^{\epsilon}_{ij})\in\mR^{n\times n}$ be the covariance matrix of $(Q^{\epsilon}_{1,\tau}, \cdots, Q^{\epsilon}_{n,\tau})^\top$.
	After algebraic simplification, under the null hypothesis of $H_0$,
	we have that
	\[\left\{
	\begin{aligned}
		\sigma^{\epsilon}_{ij}&=\frac{\{\frac{1}{n-n_{g_{i}}}+\frac{1}{n_{g_{i}-1}}\}
			p^{\epsilon}_{12}-\frac{3p^{\epsilon}_{12}-p^{\epsilon}_{22}}{(n_{g_{i}}-1)^2}}{(\frac{1}{n-n_{g_{i}}}+\frac{1}{n_{g_{i}}-1})(p^{\epsilon}_{22}-p^{\epsilon}_{21})}, \mbox{~if $i$ and $j$ belong to the same sample;~}\\
		\sigma^{\epsilon}_{ij}&=\frac{(n_{g_{i}}-1)^{1/2}(n_{g_{j}}-1)^{1/2}\{(p^{\epsilon}_{22}-(n+2)p^{\epsilon}_{12})\}}{(n-n_{g_{i}})^{1/2}(n-n_{g_{j}})^{1/2}(n-1)(p^{\epsilon}_{22}-p^{\epsilon}_{21})}, \mbox{~otherwise.~}\\
	\end{aligned}
	\right.
	\]
	Let $\hat\Sigma^{\hat\epsilon}$ be the sample counterpart of $\Sigma^{\epsilon}$ and replace $p^{\epsilon}_0$, $p^{\epsilon}_{12}$, $p^{\epsilon}_{22}$ and $\delta^{\epsilon}_{ik}$ in $\Sigma^{\epsilon}$ with
	$\hat p^{\hat\epsilon}_0=\{n(n-1)\}^{-1}\sum_{i=1}^n\sum_{k\not=i}I(\|\hat\epsilon_i-\hat\epsilon_k\|\leq\tau)$, $\hat p^{\hat\epsilon}_{12}=\{n(n-1)(n-2)\}^{-1}\sum_{i=1}^n\sum_{k\not=i}\sum_{j\not=i\not=k}\hat\delta^{\hat\epsilon}_{ik}\hat\delta^{\hat\epsilon}_{ij}$,
	$\hat p^{\hat\epsilon}_{22}=\hat p^{\hat\epsilon}_0(1-\hat p^{\hat\epsilon}_0)$
	and $\hat\delta^{\hat\epsilon}_{ik}=I(\|\hat\epsilon_i-\hat\epsilon_k\|\leq\tau)-\hat p_0^{\epsilon}$, respectively.
	It can be shown that $\hat\Sigma^{\hat\epsilon}$ is the consistent estimator of $\Sigma^{\epsilon}$.
	
	Subsequently, we follow (\ref{test_unreg}) and (\ref{adjust_test}) and construct
	the following two test statistics:
	\beq \label{test_reg}T^{\hat\epsilon}_{\tau}=\max_{1\le i\le n} \frac{(\hat{p}_i^{\hat\epsilon,bet}-\hat{p}^{\hat\epsilon, in}_i)^2}{(\frac{1}{n-n_{g_i}}+\frac{1}{n_{g_i}-1})(\hat{p}^{\hat\epsilon, (i)}_0(1-\hat{p}^{\hat\epsilon, (i)}_0)-\hat{p}^{\hat\epsilon,(i)}_{12})}\eeq
	\[\mbox{~and~} T^{\hat\epsilon,adj}_{\tau}=\max_{1\le i\le n} \big(\mM^{\hat\epsilon, adj}_{i,\tau}\big)^2,\]
	where $\mM^{\hat\epsilon, adj}_{\tau}=(\hat\Sigma^{\hat\epsilon})^{-1/2}(T^{\hat\epsilon}_{1,\tau}, \cdots, T^{\hat\epsilon}_{n,\tau})^\top=(\mM^{\hat\epsilon, adj}_{1,\tau}, \cdots, \mM^{\hat\epsilon, adj}_{n,\tau})^\top$.
	With a slight abuse of notation, we  name the above two tests MOD and CA-MOD hereafter.

	To establish the asymptotic distribution of the MOD and CA-MOD tests, we introduce the following conditions:
	\begin{itemize}
		\item[(C2)] (i) If  the $W^{(k)}$ are stochastic regressors,
we assume that $\max_{1\leq k\leq K} \|n_k^{-1}W^{(k)\top} W^{(k)}-\Sigma^{(k)}_W\|_2\to_p 0$ as $n_k\to \infty$ for any $k=1,\cdots, K$, where the
		$\Sigma^{(k)}_W$s are positive definite and  have bounded eigenvalues for any $n_k$. (ii) If the $W^{(k)}$ are deterministic regressors, we assume that $0<c_{\min}<\inf_{n_k}\lambda_{\min}(\Sigma_{n_k}^{(k)})<\sup_{n_k}\lambda_{\min}(\Sigma_{n_k}^{(k)})<c_{\min}^{-1}$ for some finite positive constant $c_{\min}$ and $\Sigma_{n_k}^{(k)}=n_k^{-1}W^{(k)\top} W^{(k)}$ for any $k=1,\cdots,K$. In addition, we assume that $\max_{1\le k\le K}\max_{1\le i\le n_k} ||W_i^{(k)}||^4$ has a finite upper bound uniformly for all $i$ and $k$.
		\item[(C3)] Assume that the $\epsilon_i^{(k)}$s are iid sub-Gaussian random variables for $i=1, \cdots, n_k$ and $k=1, \cdots, K$.
		In other words, there exist two finite positive constants $\eta$ and $C_{\eta}$
		such that $E\{\exp(\eta\epsilon_{i}^{(k)2})\}<C_{\eta}$.

	\end{itemize}
	Note that Condition (C2) is a standard assumption in linear regression models to avoid multi-collinearity (see, e.g., Shao, 2003, Guo et al. 2014). Condition (C3)
	is commonly considered in extant literature (see, e.g., Li et al. 2012 and Feng et al. 2022).
	This condition is used to evaluate the difference between $\hat\epsilon_i$ and $\epsilon_i$ for any $i=1,\cdots, n$.
	Based on the above conditions and Condition (C1), we obtain the asymptotic distributions of MOD and CA-MOD, respectively, given below.
	\bet
	Assume that  Conditions (C1)--(C3) are satisfied,
	and $p\log n/n\to 0$.
	Under $H_0$, for any $0<p_0<1$ and $x\in\mR^{+}$, as $n\to \infty$,
	we have
	\[(i) P\big\{T^{\hat\epsilon}_{\tau}<x\big\}-P(\max_{1\le i\leq n} (Z_i^{\epsilon})^2<x)\rightarrow 0, \]
	where $Z^{\epsilon}=(Z_1^{\epsilon}, \cdots, Z^{\epsilon}_n)^\top$ is a multivariate normally distributed
	random variable with mean zero and covariance matrix $\Sigma^{\epsilon}$.
	In addition, if we further assume that $\Sigma^{\epsilon}$ is positive definite for any $n$, we have
	\[
	(ii) P\Big\{T^{\hat\epsilon, adj}_{\tau}-2\log(n)+\log\log(n)\le x\Big\}\to \exp\Big\{-\frac{1}{\sqrt{\pi}}\exp(-x/2)\Big\}.
	\]
	\eet
	Theorem 5 indicates that MOD and CA-MOD are applicable to the regression setting under some proper conditions.
	The condition, $p\log n/n\to 0$, is required due to the errors induced by estimating regression models.
	To make this theorem practically useful, one can replace $\Sigma^{\epsilon}$ by its consistent
	estimator $\hat\Sigma^{\hat\epsilon}$.
	The finite sample performance of
	these two tests are given in the next section via Monte Carlo studies.
	In addition, we can apply similar techniques to those used in the proofs of
	Theorems 2 and 4 to show that MOD and CA-MOD are consistent under the alternative hypothesis in a regression setting.

	\section{Numerical studies}
	
	\subsection{Simulation Studies}
	
	In this subsection, we conduct simulation studies in two different settings.
	The first setting consists of two scenarios, $K=2$ and $K=6$. This setting is slightly modified from Chen et al. (2018) and Mukhopadhyay and Wang (2020).
	In each scenario, we consider three cases: mean shift alternatives, covariance shift alternatives, and
	distribution shift alternatives. The second setting is basically the same as the first one but applied to the multivariate regression framework. For each setting, all simulations are conducted via 1,000 realizations with nominal level $\alpha=0.05$. Thus, a well-behaved test should
	have a size around 0.05, whereas  tests with values over 0.065 are  anti-conservative and  cannot be recommended.
	As explained in Remark 2, we select $\tau$ to be the median of $\|X_i-X_j\|$s for $i,j=1,\cdots, n$ in our simulation studies.

	\begin{center}
		{\it 4.1.1. Setting IA: $K$-Sample Comparison with $K=2$}
	\end{center}
	
	{\it Case 1} (Mean shift). This example is modified from the imbalanced classes setting of Chen et al. (2018).
	We randomly generate $n/3$ observations from the multivariate normal distribution $N(0, I_p)$ and $2n/3$ observations from the multivariate normal
	distribution $N(\mu \mathbf{1}, I_p)$, where
	$\mathbf{1}=(1, \cdots, 1)^\top$ is a vector of 1s.
	Note that $\mu=0$ is used to evaluate the empirical size, while $\mu=2/\sqrt p$ is used to compute the empirical power.

	{\it Case 2} (Covariance shift).
	We randomly generate $n/3$ observations from the multivariate normal distribution $N(0, I_p)$ and $2n/3$ observations from the multivariate normal distribution $N(0, I_p+\vartheta \textbf{1} \textbf{1}^\top)$.
	Here, $\vartheta=0$ is used to evaluate the empirical size, while $\vartheta=0.8/\sqrt p$ is used to compute the empirical power.
	
	{\it Case 3} (Distribution shift).
	We randomly generate $n/3$ observations from the multivariate normal distribution $N(0, I_p)$ and $2n/3$ observations
	from the multivariate $t_{45}$ distribution with mean zero, covariance matrix $I_p$, and 45 degrees
	of freedom. This setting allows us to assess the power of the tests when the two distributions are different.

	For each setting, we consider two different sample sizes ($n=150$ and 300) and four different values for the number of
	dimensions ($p=50, 100$, 200, 500).
	Hence, $p$ can be larger than $n$.
	All of the results are based on 1,000 realizations.
	For the sake of comparison,
	we also include the edge count test of Friedman and Rafsky (1979) (EC), the generalized edge count test of Chen and Friedman (2017) (GEC),
	the weighted edge count test of Chen et al. (2018) (WEC),
	the Graph-based LP-nonparametric test (GLP) of Mukhopadhyay and Wang (2020) and
	the multisample Mahalanobis crossmatch test (MMCM) of Mukherjee et al. (2022).
	Here, GLP was implemented via the R package ``LPKsample" with $K$-means clustering suggested by the authors; we also tried the option ``mclust" for spectral clustering, and the results are similar.
	To stabilize the clusters, we set the ``$\mbox{nstart}=20$" for
	the $K$-means in implementing GLP.
	Following the approach of Chen et al. (2018), EC, GEC and WEC are all evaluated via 15-MST.
	This is because $15$-MST yields the largest power in their study, and
	the detailed definition of $M$-MST can be found in  Chen et al. (2018).
	Detailed illustrations of the five competing methods are given in Section S.4 of the supplementary material.
	Table 1 reports the empirical sizes of the seven tests in Setting IA. It indicates that
	all of the tests control empirical sizes well across the two sample sizes and four different dimensions.

	\begin{table}[htbp!]
		\begin{center}
			\caption
			{The empirical sizes of the seven tests (EC, GEC, WEC, MMCM, GLP, MOD and CA-MOD) under three different cases (i.e., mean shift, covariance shift and distribution shift alternatives) with $K=2$.}
			\vspace{0.28 cm}
      \renewcommand{\arraystretch}{0.5} 
			\begin{tabular}{ccc|cccc}
				\hline\hline
				Setting &$n$ & Methods & $p=50$ & $p=100$ & $p=200$  &  $p=500$ \\
				\hline\hline
				Size-all cases  & 150    &EC   &    0.051 &    0.045 &    0.042 &    0.040 \\
				&        &GEC   &    0.044 &    0.049 &    0.037 &    0.050 \\
				&        &WEC   &    0.042 &    0.046 &    0.044 &    0.070 \\
				&        &MMCM  &    0.040 &    0.051 &    0.052 &   0.062  \\
				&        &GLP   &    0.035 &    0.045 &   0.026  &   0.067 \\
				&        &MOD   &    0.047 &    0.056 &    0.049 &    0.051 \\
				&        &CA-MOD  &    0.044 &    0.040 &    0.044 &    0.046 \\
				\hline
				& 300    &EC    &    0.044 &    0.040 &    0.049 &    0.050 \\
				&        &GEC   &    0.060 &    0.048 &    0.055 &    0.045 \\
				&        &WEC   &    0.053 &    0.051 &    0.052 &    0.041 \\
				&        &MMCM  &    0.042 &    0.061 &    0.045 &   0.064  \\
				&        &GLP   &    0.045 &    0.050 &   0.036  &   0.048 \\
				&        &MOD   &    0.047 &    0.037 &    0.059 &    0.045 \\
				&        &CA-MOD  &    0.039 &    0.046 &    0.046 &    0.050 \\
				\hline\hline
			\end{tabular}
		\end{center}
	\end{table}

	\begin{table}[htbp!]
		\begin{center}
			\caption
			{The empirical powers of the seven tests (EC, GEC, WEC, MMCM, GLP, MOD and CA-MOD) under three different cases (i.e., mean shift, covariance shift and distribution shift alternatives) with $K=2$.}
			\vspace{0.28 cm}
   \renewcommand{\arraystretch}{0.5} 
   \begin{tabular}{ccc|cccc}
				\hline\hline
				Setting &$n$ & Methods & $p=50$ & $p=100$ & $p=200$  &  $p=500$ \\
				\hline\hline
				Case 1   & 150       &EC    &    0.786 &    0.523 &    0.325 &    0.158 \\
				&           &GEC   &    0.984 &    0.897 &    0.685 &    0.393 \\
				&           &WEC   &    0.993 &    0.939 &    0.772 &    0.478 \\
				&           &MMCM  &    0.746 &    0.585 &    0.361 &    0.190 \\
				&           &GLP   &    0.874 &    0.626 &    0.332 &    0.123 \\
				&           &MOD   &    0.959 &    0.679 &    0.376 &    0.180 \\
				&           &CA-MOD  &    0.949 &    0.656 &    0.362 &    0.193 \\
				\hline
				& 300       &EC    &    0.957 &    0.797 &    0.532 &    0.273 \\
				&           &GEC   &    1.000 &    0.995 &    0.959 &    0.698 \\
				&           &WEC   &    1.000 &    0.999 &    0.984 &    0.775 \\
				&           &MMCM  &    0.961  &    0.865 &    0.612 &    0.334 \\
				&           &GLP   &   0.995   &    0.951 &    0.753 &    0.322 \\
				&           &MOD   &    1.000 &    0.990 &    0.793 &    0.337 \\
				&           &CA-MOD  &    1.000 &    0.991 &    0.787 &    0.339 \\
				\hline\hline
				Case 2    & 150       &EC   &    0.037 &    0.024 &    0.033 &    0.023 \\
				&           &GEC   &    0.469 &    0.492 &    0.491 &    0.518 \\
				&           &WEC   &    0.397 &    0.415 &    0.430 &    0.449 \\
				&           &MMCM  &    0.202 &     0.213 &    0.234 & 0.296 \\
				&           &GLP   &    0.155 &     0.177 &    0.301 & 0.392 \\
				&           &MOD   &    0.379 &    0.392 &    0.425 &    0.427 \\
				&           &CA-MOD  &    0.644 &    0.663 &    0.683 &    0.694 \\
				\hline
				& 300       &EC    &    0.041 &    0.032 &    0.025 &    0.035 \\
				&           &GEC   &    0.703 &    0.746 &    0.755 &    0.739 \\
				&           &WEC   &    0.641 &    0.703 &    0.729 &    0.711 \\
				&           &MMCM  &    0.321 &    0.362 &    0.365 &    0.413 \\
				&           &GLP   &    0.186 &     0.274 &    0.373 &   0.556 \\
				&           &MOD   &    0.844 &    0.851 &    0.866 &    0.874 \\
				&           &CA-MOD  &    0.976 &    0.984 &    0.980 &    0.989 \\
				\hline\hline
				Case 3   & 150       &EC    &    0.067 &    0.133 &    0.221 &    0.398 \\
				&           &GEC   &    0.060 &    0.138 &    0.531 &    0.997 \\
				&           &WEC   &    0.057 &    0.094 &    0.329 &    0.613 \\
				&           &MMCM  &    0.045 &    0.071 &    0.086 &   0.092 \\
				&           &GLP   &   0.032  &    0.077 &    0.079 &    0.146 \\
				&           &MOD   &    0.065 &    0.178 &    0.548 &    0.985 \\
				&           &CA-MOD  &    0.203 &    0.492 &    0.937 &    1.000 \\
				\hline
				& 300       &EC    &    0.179 &    0.507 &    0.832 &    0.977 \\
				&           &GEC   &    0.111 &    0.425 &    0.925 &    1.000 \\
				&           &WEC   &    0.062 &    0.134 &    0.345 &    0.733 \\
				&           & MMCM &  0.095   &  0.051  &      0.071  &  0.062 \\
				&           & GLP  & 0.076   &   0.043  &   0.132   &  0.244 \\
				&           &MOD   &    0.194 &    0.653 &    0.990 &    1.000 \\
				&           &CA-MOD  &    0.350 &    0.897 &    1.000 &    1.000 \\
				\hline\hline
			\end{tabular}
		\end{center}
	\end{table}
	
	In power comparisons,
	Table 2 shows that WEC performs the best with mean shift alternatives.
	This finding is expected, since WEC was designed to enhance the power of EC for the location shift case; see Chen et al. (2018) for details.
	In addition,
	we also find that CA-MOD and MOD are
	comparable to EC, GLP and MMCM, and they are
	inferior to GEC and WEC.
	It is of interest to note that all powers decrease as $p$ gets large, when the mean shift alternatives are $\mu=1.3/\sqrt p$, for $p=50,100, 200$ and $500$.  This is because the average signal of any component in $\mu$ is only $1.3/\sqrt{p}$, and it decreases to 0 as $p$ increases, although $\|\mu\|=1.3$ remains constant.
	As a result, it becomes more difficult to differentiate the null and alternative hypotheses, which leads to our finding.

	As for the covariance shift case, Table 2 indicates that CA-MOD performs the best. Specifically, it is better than GEC, WEC and MOD, and outperforms EC, GLP and MMCM.
	It is not surprising that
	all powers increase as $p$ becomes large since this is due to the design of the  variance shift alternatives, $\vartheta=1/\sqrt p$, for $p=50,100, 200$ and $500$.
	Finally, in Case 3 with the distribution shift alternatives, CA-MOD also performs the best.
	In addition, both MOD and CA-MOD are
	better than GEC, WEC and EC, and outperform GLP and MMCM significantly. Moreover, it is expected that all powers get large as $p$ increases,
	and this is due to the design of the alternative with the multivariate $t_{45}$ distribution.

	In addition to testing the normal distribution versus the $t$-type distribution, we  conduct a simulation for testing the normal distribution against the mixture normal distribution
	to examine the impact of outliers on seven tests. Specifically, we randomly generate $n/3$ observations from the multivariate normal distribution $N(0, I_p)$ and $2n/3$ observations
	from the mixture normal distribution $(1-\eta)N(0, I_p)+\eta N(20, 3 I_p)$. We set $\eta=0.05$, and
	find that the empirical powers of these seven tests are comparable and all of them approach to 1. To save space, we do not
	report them here.

	\begin{center}
		{\it 4.1.2. Setting IB: $K$-Sample Comparison with $K=6$}
	\end{center}
	
	In this study, we consider $K=6$ with total sample sizes $n=300$ and 600, four different numbers of dimensions ($p=25, 50$, 100, 200)
	and the dimensions of variables $X_i^{(k)}$ are the same as those in Setting IA.
	Because EC, GEC and WEC are not
	applicable for $K>2$, we only compare MOD and CA-MOD
	with the GLP test of Mukhopadhyay and Wang (2020) and MMCM test of Mukherjee et al. (2022).
	
	{\it Case 1} (Mean shift).
	For $k=1, \cdots, K/2$,
	we randomly generate $X_i^{(k)}$ with $i=1,\cdots, n/K$ observations from the
	multivariate normal distribution $N(0, I_p)$. The observations in the remaining $K/2$ samples are simulated from the multivariate normal distribution $N(\mu \mathbf{1}, I_p)$.
	Note that $\mu=0$ is used to evaluate the empirical size, while $\mu=2.6/\sqrt p$ is used to compute the empirical power.

	{\it Case 2} (Covariance shift).
	For $k=1, \cdots, K/2$,
	we randomly generate $X_i^{(k)}$ with $i=1,\cdots, n/K$ observations from the
	multivariate normal distribution $N(0, I_p)$. The observations of in the remaining $K/2$ samples are simulated from the multivariate normal distribution $N(0, I_p+\vartheta\textbf{1}\textbf{1}^\top)$,
	where $\textbf{1}$ is a vector of 1s.
	The $\vartheta=0$ is used to evaluate the empirical size, while $\vartheta=1/\sqrt p$ is used to calculate the empirical power.

	{\it Case 3} (Distribution shift).
	For $k=1, \cdots, K/2$,
	we randomly generate $X_i^{(k)}$ with $i=1,\cdots, n/K$ observations from the
	multivariate normal distribution $N(0, I_p)$. The observations in the remaining $K/2$ samples are simulated
	from the multivariate $t_{30}$ distribution with the covariance matrix $I_p$ and 30 degrees of freedom. This setting allows us
	to assess the power of these tests when the $K$-sample distributions are different.
	
	\begin{table}[htbp!]
		\begin{center}
			\caption
			{The empirical sizes and powers of the four tests (MMCM, GLP, MOD and CA-MOD) under three different cases (i.e., mean shift, covariance shift and distribution shift alternatives) with $K=6$.}
			\vspace{0.28 cm}
      \renewcommand{\arraystretch}{0.5} 
			\begin{tabular}{ccc|cccc}
				\hline\hline
				Setting & $n$ &  Methods & $p=50$ & $p=100$ & $p=200$  &  $p=500$ \\
				\hline
				Sizes  &    300       &MMCM     &  0.040 & 0.040 & 0.040 & 0.030 \\
				&              & GLP     &    0.076  &  0.054  & 0.096 &  0.081 \\
				&              &MOD      &    0.042 &    0.036 &    0.037 &    0.039 \\
				&              &CA-MOD   &    0.056 &    0.041 &    0.034 &    0.047 \\
				\hline
				&  600         & MMCM    &   0.065 & 0.067 & 0.060 & 0.055 \\
				&              & GLP     & 0.077  & 0.072  &  0.073 &  0.042 \\
				&              &MOD      &    0.043 &    0.039 &    0.049 &    0.038 \\
				&              &CA-MAD   &    0.051 &    0.037 &    0.043 &    0.048 \\
				\hline\hline
				Power-Case 1 &   300   &MMCM     &    0.980 &  0.870 & 0.520 & 0.235 \\
				&         & GLP     &    1.000 &  1.000 & 1.000 &  0.930 \\
				&         &MOD      &    0.948 &    0.608 &    0.368 &    0.209 \\
				&         &CA-MAD   &    0.940 &    0.601 &    0.369 &    0.217 \\
				\hline
				&   600   &MMCM     &  1.000 &  1.000 & 0.990  &  0.610 \\
				&         & GLP     &   1.000 & 1.000 & 1.000 & 1.000 \\
				&         &MOD      &    1.000 &    0.999 &    0.823 &    0.523 \\
				&         &CA-MOD   &    1.000 &    0.997 &    0.855 &    0.538 \\
				\hline\hline
				Power-Case 2 &    300  & MMCM    &   0.192 &  0.236 &  0.243 &  0.255 \\
				&         & GLP     &   0.492 &  0.543 &  0.561 &  0.577  \\
				&         &MOD      &    0.401 &    0.421 &    0.436 &    0.442 \\
				&         &CA-MOD   &    0.574 &    0.586 &    0.592 &    0.604 \\
				\hline
				&    600 & MMCM    &  0.410   & 0.413   & 0.562  &  0.651 \\
				&        & GLP     &   0.921  &  0.942  &  0.957 & 0.971 \\
				&        & MOD     &    0.801 &    0.817 &    0.839 &    0.842 \\
				&        &CA-MOD   &    0.963 &    0.965 &    0.968 &    0.972 \\
				\hline\hline
				Power-Case 3 &   300   & MMCM    &   0.025 &  0.065  & 0.040 & 0.035 \\
				&         & GLP     &   0.075 & 0.130   & 0.345 & 0.615 \\
				&         & MOD     &    0.166 &    0.410 &    0.706 &    0.955 \\
				&         &CA-MOD   &    0.255 &    0.656 &    0.971 &    1.000 \\
				\hline
				&600     & MMCM    &  0.045   &  0.055  & 0.050  & 0.060 \\
				&        & GLP     & 0.090    &  0.270   &  0.580   & 0.940 \\
				&        &MOD      &    0.377 &    0.811 &    0.995 &    1.000 \\
				&        &CA-MOD   &    0.509 &    0.944 &    1.000 &    1.000 \\
				\hline\hline
			\end{tabular}
		\end{center}
	\end{table}

	Table 3 shows that MOD, CA-MOD and MMCM control empirical sizes well across all $n$ and $p$. However,
	GLP can yield size distortions;  this is because the
	GLP test involves a spectral clustering procedure, and its testing results rely on the accuracy of clustering.
	Thus, this method can be unstable for a relatively large $K=6$, and we do not discuss it in power comparisons.
	As for the power performance,
	all three test statistics,  MOD, CA-MOD and MMCM, tend to 1 or get large as $n$ increases.
	In Case 1 with the mean shift alternatives, MOD and CA-MOD are inferior to (or comparable to) MMCM.
	In both Case 2  with the covariance shift alternatives
	and Case 3 with the distribution shift alternatives,  CA-MOD and MOD are superior to MMCM, and CA-MOD performs the best.
	In sum, Table 3 demonstrates that the finite sample performance of MOD and CA-MOD is  satisfactory
	based on the corresponding  asymptotic results obtained in Theorems 1 and 2.

	Based on the discussion below Theorem 2,
	we choose $\tau$ to be the median of the $\|X_i-X_j\|$s in previous studies.
	As mentioned in Remark 2,  we also conduct simulation studies with two additional $\tau$s, namely the  25\% and 75\% quantiles of the $\|X_i-X_j\|$s.  The results show that the median of the $\|X_i-X_j\|$s performs the best in most cases; see Tables S.1--S.4 of the supplementary material.

	\begin{center}
		{\it 4.1.3. Setting II: $K$-Sample Comparison in Multivariate Regression}
	\end{center}

	To study the $K$-sample comparison under a multivariate regression setting,
	we consider four different values for the number of dimensions ($p=25, 50$, 100, 200). In addition, the
	sample sizes  for $K=2$  are  $n=$150 and 300, and the
	sample sizes for  $K=6$ are
	$n=600$ and 900.
	The data are generated from model (\ref{reg_model}), where
	each element of the covariate $W_i^{(k)}\in\mR^d$ is independent
	and identically generated from a standard normal distribution,
	the regression coefficients are $\beta^{(k)}=(\beta_{jl}^{(k)})\in\mR^{p\times d}$ with $\beta_{jl}^{(k)}=1$ for all $k$,
	the number of covariates is
	$d=2$, $i=1,\cdots,n$, $j=1,\cdots,p$, $l=1,\cdots,d$, and $k=1,\cdots, K$.
	In addition, the random noise $\epsilon_i^{(k)}$s are generated according to
	those observations in Setting IA with $K=2$ and Setting IB with $K=6$.
	For $K=2$, the alternatives settings are the same as those in Section 4.1.1.
	For $K=6$, we slightly modify the setting in Section 4.1.2, and let $\mu=1.8/\sqrt{p}$
	for mean shift alternative and $\nu=0.75/\sqrt{p}$ for the variance shift.
	\begin{table}[htbp!]
		\begin{center}
			\caption
			{The empirical sizes of the seven tests (EC, GEC, WEC, GLP, MMCM, MOD and CA-MOD) under three different cases (i.e., mean shift, covariance shift and distribution shift alternatives) when $K=2$ in a regression setting.}
      \renewcommand{\arraystretch}{0.5} 
			\vspace{0.28 cm}
			\begin{tabular}{cc|cccc}
				\hline\hline
				$n$ & Methods & $p=25$ & $p=50$ & $p=100$  &  $p=200$ \\
				\hline\hline
				150      &EC    &    0.032 &    0.052 &    0.041 &    0.056 \\
				&GEC   &    0.391 &    0.441 &    0.524 &    0.736 \\
				&WEC   &    0.068 &    0.073 &    0.128 &    0.268 \\
				&MMCM  &    0.072 &    0.116 &    0.117 &    0.204 \\
				&GLP   &    0.052 &    0.067 &    0.077 &    0.095 \\
				&MOD   &    0.045 &    0.056 &    0.062 &    0.064 \\
				&CA-MOD  &    0.044 &    0.052 &    0.058 &    0.062 \\
				\hline
				300      &EC    &    0.053 &    0.053 &    0.051 &    0.053 \\
				&GEC   &    0.367 &    0.387 &    0.418 &    0.521 \\
				&WEC   &    0.059 &    0.047 &    0.065 &    0.114 \\
				&MMCM  &    0.055 &    0.045 &    0.071 &   0.077 \\
				&GLP   &    0.052 &    0.047 &    0.052 &   0.046 \\
				&MOD   &    0.045 &    0.049 &    0.048 &    0.058 \\
				&CA-MOD  &    0.049 &    0.037 &    0.052 &    0.045 \\
				\hline\hline
			\end{tabular}
		\end{center}
	\end{table}

	Tables 4 reports the results of the seven tests with $K=2$.
	It indicates that both MOD and CA-MOD control size well across all cases.
	We next consider the sizes of the other five tests,
	although EC, GEC, WEC, MMCM and GLP are
	not designed primarily for the regression setting.
	Table 4 shows that GEC   does not control its size well in the regression
	setting; hence, we do not discuss it in power comparisons.
	In addition, our results indicate that the EC and GLP tests control size well, except that
	GLP has size inflation when $n=150$ and  $p=100$,
	and when $n=150$ and $p=200$.
	Moreover, WEC and MMCM yield size inflation when $n=150$ or $p=200$.
	Thus, we do not discuss their corresponding powers under these  settings
	with size inflation.
	To save space, we present power comparisons in Table S.6 of Section S.5 in the supplementary material.

	For the mean shift alternatives,
	Table S.6 shows that both
	MOD and CA-MOD outperform EC and
	GLP. Note that  we do not discuss the power of  GLP with $p=200$ and $n=150$  due to its size inflation.
	In addition, MOD and CA-MOD outperform or perform comparably to
	WEC and MMCM when  $p=200$ or $n=150$ is excluded from power comparisons.
	Finally, in testing covariances and distributions,  MOD and CA-MOD outperform EC, WEC, MMCM and GLP in all settings. Overall, CA-MOD performs the best.

	For the setting with $K=6$, Table S.7 in the supplementary material reports the sizes and powers of MMCM, GLP, MOD and CA-MOD.
	Note that we exclude three tests,  EC, GEC and WEC, since they are not applicable for  $K>2$. Table S.7 indicates that both MOD and CA-MOD control sizes well across all cases.
	However, MMCM yields size distortions for $n=600$ and $p\ne 25$, and GLP has
	size distortions when $p=100$ and 200.
	In comparing the powers of these four tests, Table S.7 shows similar findings qualitatively to those in Table 4.
	In sum,  both MOD and CA-MOD tests perform satisfactorily, and CA-MOD performs the best in general.

	\subsection{Real Data Analysis}

		Since the   efficient-markets theory was introduced by Nobel Laureate Eugene Fama (1970), testing the efficiency of the stock market  has become a central topic in finance.
		When a market is efficient, the day-of-the-week effect no
		longer exists (see, e.g.,  Gibbons and Hess, 1981).
		This  motivates us to employ the two proposed methods to test the day-of-the-week effect in the Chinese stock market and assess market efficiency (see, e.g., Kohers et al. 2004 and Dicle and Levendis, 2014).
		One possible approach is to test whether the distributions
		of stock returns recorded on Monday, Tuesday, Wednesday, Thursday, and Friday are all equal.
		To this end, we collect the stock data from the
		WIND database, which is one of the most authoritative financial databases in China. After deleting the stocks with missing values, we obtain data
		that consists of 48 Monday records, 49 Tuesday records, 50 Wednesday records, 49 Thursday records, and 47 Friday records, with  a total of 898 stocks from 01/02/2020 to 12/31/2020.
		
	The goal of this study is testing the equality of distributions of stock returns recorded on Monday, Tuesday, Wednesday, Thursday, and Friday.
		Since $K=5$, we only consider
		our proposed MOD and CA-MOD tests since the GLP test is unstable and the MMCM test is less  powerful than ours  based on simulation results with $K=6$.
		The resulting $p$-values of the MOD and CA-MOD tests  are 0.199 and 0.247, respectively.
		Neither of them rejects the null hypothesis of equal distribution with nominal level 5\%. Accordingly, there is no sufficient evidence to
		support the existence of the day-of-the-week effect.

	It is worth noting that the individual stock return can be influenced by the market return based on the seminal Capital Asset Pricing Model (Sharpe, 1964). Hence, it is of interest to
	test the day-of-the-week effect after adjusting for the market return.
	To this end,
	we use the Hushen 300 index (an index representing the performance of the 300 largest stocks on the Chinese stock market) as the market index,
	and regress the stock returns on the market index separately on Monday, Tuesday, Wednesday, Thursday, and Friday records. Subsequently,
	we conduct the MOD and CA-MOD tests based on the residuals obtained from fitting the regression model (5). The resulting $p$-values for these two tests are
	0.140 and 0.151, respectively.
	Both tests
	fail to reject the null hypothesis of equal distributions at the nominal level of 5\%.
	The above results indicate that
	there is no evidence to
	support the existence of the day-of-the-week effect with or without adjusting for the market performance.
	Hence, we conclude that
	the Chinese stock market is becoming more efficient after the structural reform during 2005--2006 (see, e.g., Chong et al. 2012 and Carpenter et al. 2021).

	\section{Concluding remarks}
	In this paper, we propose a novel maximum-of-differences (MOD) test to compare $K$-sample distributions. Specifically, we evaluate the
	pairwise distance between pooled observations from $K$ samples, and  then construct the test by
	maximizing the differences of the``within" and the ``between" probabilities from all observations. In addition, we
	obtain the covariance-adjusted version of the MOD test, CA-MOD.
	To broaden their usefulness, we also extend  both tests to the multivariate regression setting.

		To conclude the article, we identify four possible avenues for future research. First, adapt Feng et al.'s (2022) approach by combining  the maximum-type test and the sum-type test together to obtain a  powerful test under dense alternatives, such as  $H_1: F_k\not=F_l$ for the most of $k\not=l$. Second, in addition to  the $L_2$ distance considered in the paper, it is of interest to study the optimal choice of the distance measure to form test statistics. Third, construct the test statistic with a varying $\tau$. Finally, take into account correlated observations to construct the test (see, e.g., Xue and Yao, 2020).
	
	\section*{Supplementary material}


The supplementary material includes five sections.
Section S.1 presents a lemma that is useful for proving Theorems 1--5.
Sections S.2 provides the proofs of Theorems.
Section S.3 shows the detailed derivation of $\Sigma=(\sigma_{ij})$.
Section S.4 introduces the five competing methods, which are used in the simulation section of the main paper.
Section S.5 includes three parts, which report additional simulation results in Tables S.1--S.7.
Specifically, Part I conducts simulation studies with different $\tau$s,
	Part II provides simulation findings with small $p$s (i.e., $p=2$, 5 and 10), and Part III reports simulation results in regression settings when $K=2$ and 6.
In the remainder of this supplementary material, we denote $C_b^{\phi}(\mathbb{R})$ as the set of
functions defined on $\mR$ having $\phi$-th continuous and bounded derivatives, for any positive integer $\phi$.

\section*{Section S.1:  Lemma 1}

In order to employ Lemma J.2 of \cite{c2013} for showing
Theorems 1--5, we rewrite it given below.

\begin{la}\label{le1}
	Let $V=(V_1,\cdots, V_{\nu})^\top\in \mathbb{R}^{\nu}$ and $U=(U_1,\cdots, U_{\nu})^\top\in \mathbb{R}^{\nu}$ be centered Gaussian random vectors  with mean zero
	and covariance
	matrices $\Sigma^{V}=(\Sigma_{j_1j_2}^{V})$ and $\Sigma^{U}=(\Sigma_{j_1j_2}^{U})$,
	respectively. In addition, let $\Delta_{0}=\max _{1 \leqslant j_1, j_2 \leqslant p} \mid \Sigma_{j_1j_2}^{V}-\Sigma_{j_1j_2}^{U} \mid$.
	Suppose that there exist two finite positive constants $c_{1}<C_{1}$
	such that $\bar{\sigma}:=$ $\max _{1 \leqslant j \leqslant \nu} \mathrm{E}\left[U_{j}^{2}\right] \leqslant C_{1}$ for all $1 \leqslant j \leqslant \nu$ and
	$b_{\nu}:=\mathrm{E}\left[\max _{1 \leqslant j \leqslant \nu} U_{j}\right] \geqslant$
	$c_{1} \sqrt{\log p}.$ Then, there exist finite constants $c_2>0$ and $C_2>0$ such that, depending only on $c_{1}$ and $C_{1}$,
	\[
	\sup _{t \in \mathbb{R}}\left|\mathrm{P}\left(\max _{1 \leqslant j \leqslant \nu} V_{j} \leqslant t\right)-\mathrm{P}\left(\max _{1 \leqslant j \leqslant \nu} U_{j} \leqslant t\right)\right|
	\]
	\[ \leqslant C_2 \Delta_{0}^{1 / 3}\left(1 \vee \log \left(\nu / \Delta_{0}\right)\right)^{2 / 3}
	+C_2 \nu^{-c_2} \sqrt{1 \vee \log \left(\nu / \Delta_{0}\right)},
	\]
	where $a \vee b=\max\{a, b\}$ for any constants $a, b \in \mathbb{R}$.
\end{la}

\section*{Section S.2: Proof of Theorems}
\subsection*{S.2.1 Proof of Theorem 1}

Taking the definition of $Q_{i,\tau}$ in Section 2.2, we re-express it as follows:
\begin{align*}
	Q_{i,\tau}=&\frac{\hat{p}_i^{bet}-\hat{p}^{in}_i}{\{(\frac{1}{n-n_{g_i}}+\frac{1}{n_{g_i}-1})(p_{22}-p_{12})\}^{1/2}}\doteq \frac{1}{\sqrt n}\sum_{j=1}^n V_{ij,\tau},
\end{align*}
where $V_{ij,\tau}=\frac{\sqrt n}{(n-n_{g_i})\{(\frac{1}{n-n_{g_i}}+\frac{1}{n_{g_i}-1})(p_{22}-p_{12})\}^{1/2}}\big\{I(||X_i-X_j||\leq\tau)-p_0\big\}$
for $j\not\in \mathcal{C}_{g_i}$ and $V_{ij,\tau}=-\frac{\sqrt n}{(n_{g_i}-1)\{(\frac{1}{n-n_{g_i}}+\frac{1}{n_{g_i}-1})(p_{22}-p_{12})\}^{1/2}}\big\{I(||X_i-X_j||\leq\tau)-p_0\big\}$
for $j\in \mathcal{C}_{g_i}$ and $j\not=i$.
Under Condition (C1), all of the $V_{ij,\tau}$s have the same order of $O(1)$.
Then define $T_0=\max_{1\le i \le n} Q_{i,\tau}$ and $Z_0=\max_{1\le i \le n} Y_i$,
where $Y_i=\frac{1}{\sqrt{n}}\sum_{j=1}^n y_{ij}$ and $y_{ij}$ are independent and identically distributed standard normal variables,
and $Y\doteq (Y_1,\cdots,Y_n)^\top$ has the same covariance matrix as $Q\doteq(Q_{1,\tau},\cdots,Q_{n,\tau})^\top$, i.e., $\Sigma$.
We next prove the theorem via  the following two steps.

{\sc Step I}. We demonstrate that,
for any $t\in \mathbb{R}$,
\begin{align}
	\sup_{t\in \mathbb{R}}|P(T_0\le t)-P(Z_0\le t)|\to 0, \label{A.1}
\end{align}
as $n\to \infty$.
Define $F_{\kappa}(\bm z)\doteq \kappa^{-1}\log \left(\sum_{j=1}^n e^{\kappa z_j}\right)$,
where $\bm z=(z_1,\cdots,z_n)^\top \in \mathbb{R}^n$ and $\kappa$ is a finite
positive constant.
Using the fact that $e^{\max_j \kappa z_j}\leq \sum_{j=1}^n e^{\kappa z_j}\leq n e^{\max_j \kappa z_j}$, we have that,
for any $\bm z\in \mathbb{R}^n$ and $\kappa\in\mR^+$,
\begin{align}
	0\le F_{\kappa}(\bm z)-\max_{1\le i\le n}z_j\le \kappa^{-1}\log n. \label{A.2}
\end{align}
Define $e_{\kappa}:=\kappa^{-1} \log n$. Then, for any $t\in\mR$, we have
\begin{align}
	P(T_0\le t) \le P\big\{F_{\kappa}(Q)\le t+e_\kappa\big\}. \label{A.3}
\end{align}
Consider a $C_{b}^{3}(\mathbb{R})$-function $g_{0}: \mathbb{R} \rightarrow[0,1]$ such that $g_{0}(s)=1$ for $s \leqslant 0$ and $g_{0}(s)=0$ for $s \geqslant 1$. Given $t \in \mathbb{R}$, we
define $g(s)=$ $g_{0}\big\{\psi(s-t-e_{\kappa})\psi\big\}$, where $\psi=n^{1/8}$ and
$\kappa=\psi \log n$.
In addition, define $G_k\doteq \sup_{z\in \mR}|\partial^k g(z)/\partial z|$.
Then, we have
$G_{0}=1, G_{1}\lesssim \psi, G_{2} \lesssim \psi^{2}$ and $G_{3} \lesssim \psi^{3}$,
where $a\lesssim b$ means $a/b<\infty$ for any arbitrary constants $a$ and $b$.
Accordingly, by the definition of $g(\cdot)$, we have
\begin{align}
	P\big\{F_{\kappa}(Q)\le t+e_\kappa\big\}\le E\big\{g(F_\kappa(Q))\big\}. \label{A.4}
\end{align}

We next show that
\begin{align}
	E\big\{g(F_\kappa(Q))\big\}-E\big\{g(F_\kappa(Y))\big\}\le C_3\left(\psi^{3}+\kappa \psi^{2}+\kappa^{2} \psi\right)\left(n^{-1 / 2} \overline{\mathrm{E}}\left[S_{i}^{3}\right]\right)+o(1), \label{A.5}
\end{align}
for a finite positive constant $C_3$,
$S_{i}:=\max _{1 \leqslant j \leqslant n}(\left|y_{i j}\right|+|V_{ij,\tau}|)$ and $\bar{E}(s_i)=n^{-1}\sum_{i=1}^nE(s_i)$ for any variable $s_i$.
For $t \in[0,1]$, we consider the Slepian interpolation between $Y$ and $Q$ as follows:
$$
H(t):=\sqrt{t} Q+\sqrt{1-t} Y=\sum_{j=1}^{n} H_{j}(t) ~~ {\mbox{and}} ~~ H_{j}(t):=\frac{1}{\sqrt{n}}\big(\sqrt{t} V_{\cdot j,\tau}+\sqrt{1-t} y_{\cdot j}\big),
$$
where $V_{\cdot j,\tau}=(V_{1j,\tau},\cdots,V_{nj,\tau})^\top$, $y_{\cdot j}=(y_{1j},\cdots,y_{nj})^\top$ and
$H_{j}(t)=(H_{1j}(t),\cdots, H_{nj}(t))^\top\in\mR^n$.
We further employ Stein's leave-one-out expansion and have that
$$
H^{(ij)}(t):=H(t)-H_{i}(t)-H_{j}(t)=(h^{(ij)}_1(t),\cdots,h^{(ij)}_n(t))^\top.
$$
Let $m:=g \circ F_{\kappa}$ and define $\Psi(t)=\mathrm{E}\big\{m(H(t))\big\}$ for $m:=g \circ F_{\kappa}$.
Then, we have $E\big\{g(F_\kappa(Q))\big\}-E\big\{g(F_\kappa(Y))\big\}=\Psi(1)-\Psi(0).$
In addition, for any function $f: \mathbb{R}^{\varsigma} \rightarrow \mathbb{R}$, denote a generic notation
$\partial_{i} f(u)=\partial f(u) / \partial u_{i}$ for any $i=1, \cdots, \varsigma$, where $u=\left(u_{1}, \cdots, u_{\varsigma}\right)^{\top}$.
By  Taylor's theorem (see H\"ormander, 1976, pp 12-13), we have that
\begin{align*}
	\partial_{i} m(H(t))=&\partial_{i} m\left(H^{(ij)}(t)\right)+\sum_{k=1}^n\partial_{i} \partial_{k} m\left(H^{(ij)}(t)\right) \{H_{kj }(t)+H_{ki}(t)\}\\
	&+\sum_{k,l=1}^n \int_{0}^{1}(1-\xi)\partial_{i} \partial_{k} \partial_{l} m\left(H^{(ij)}(t)+\xi H_{j}(t)+\xi H_{i}(t)\right)\\
	&\times\{H_{kj}(t) H_{lj}(t)+ H_{ki}(t) H_{li}(t)+H_{kj}(t) H_{li}(t)+ H_{ki}(t) H_{lj}(t)\}d \xi.
\end{align*}
This leads to
\begin{align*}
	\mathrm{E}[m(Q)-m(Y)] &=\Psi(1)-\Psi(0)=\int_{0}^{1} \Psi^{\prime}(t) d t \\
	&=\frac{1}{2} \sum_{i=1}^{n}\sum_{j=1}^{n}  \int_{0}^{1} \mathrm{E}\left[\partial_{i} m(H(t)) \dot{H}_{i j}(t)\right] d t\doteq
	\frac{1}{2}(\Delta_1+\Delta_2+\Delta_3),
\end{align*}
where $\dot{H}_{i j}(t)=\frac{d}{d t} H_{i j}(t)=\frac{1}{\sqrt{n}}\big(\frac{1}{\sqrt{t}} V_{i j,\tau}-\frac{1}{\sqrt{1-t}} y_{i j}\big)$,
\begin{align*}
	\Delta_1 &= \sum_{i=1}^{n} \sum_{j=1}^{n}\int_{0}^{1} \mathrm{E}\left[\partial_{i} m\left(H^{(ij)}(t)\right) \dot{H}_{i j}(t)\right] d t,\\
	\Delta_2 &=\sum_{i, k=1}^{n} \sum_{j=1}^{n} \int_{0}^{1} \mathrm{E}\left[\partial_{i} \partial_{k} m\left(H^{(ij)}(t)\right) \dot{H}_{i j}(t) \{H_{kj }(t)+H_{ki}(t)\}\right] d t ~~ {\mbox{and}} \\
	\Delta_3 &=\sum_{i, k, l=1}^{n} \sum_{j=1}^{n} \int_{0}^{1} \int_{0}^{1}(1-\xi) \mathrm{E}\Bigg[\partial_{i} \partial_{k} \partial_{l} m\left(H^{(ij)}(t)+\xi H_{j}(t)+\xi H_{i}(t)\right) \\
	&\times\dot{H}_{i j}(t) \{H_{kj}(t) H_{lj}(t)+ H_{ki}(t) H_{li}(t)+H_{kj}(t) H_{li}(t)+ H_{ki}(t) H_{lj}(t)\}\Bigg] d \xi d t.
\end{align*}

We study the above three parts separately.
Define $e_k^{(ij)}=(0,\cdots,0,h^{(ij)}_k(t),0,\cdots,0)^\top$ and $\tilde{H}^{(ij)}(t)=H^{(ij)}(t)-e_i^{(ij)}-e_j^{(ij)}$.
Then, by  Taylor's Theorem, we have
\begin{align*}
	\partial_{i} m\left(H^{(ij)}(t)\right)=&\partial_{i} m\left(\tilde H^{(ij)}(t)\right)+\int_0^1(1-\xi)\partial_{i}^2 m\left(\tilde H^{(ij)}(t)+\xi e_i^{(ij)}(t)+\xi e_j^{(ij)}(t)\right) z_i^{(ij)}(t)d \xi\\
	&+\int_0^1(1-\xi)\partial_{i}\partial_j m\left(\tilde H^{(ij)}(t)+\xi e_i^{(ij)}(t)+\xi e_j^{(ij)}(t)\right) h_j^{(ij)}(t)d \xi.
\end{align*}
Accordingly, we obtain
\begin{align*}
	\Delta_1 &= \sum_{i=1}^{n} \sum_{j=1}^{n}\int_{0}^{1} \mathrm{E}\left[\partial_{i} m\left(H^{(ij)}(t)\right) \dot{H}_{i j}(t)\right] d t \\
	&= \sum_{i=1}^{n} \sum_{j=1}^{n}\int_{0}^{1} \mathrm{E}\left[\partial_{i} m\left(\tilde H^{(ij)}(t)\right) \dot{H}_{i j}(t)\right] d t\\
	&+\sum_{i=1}^{n} \sum_{j=1}^{n}\int_{0}^{1} \int_{0}^{1}(1-\xi)\mathrm{E}\left[\partial_{i}^2 m\left(\tilde H^{(ij)}(t)+\xi e_i^{(ij)}(t)+\xi e_j^{(ij)}(t)\right) \dot{H}_{i j}(t)h_i^{(ij)}(t)\right] d \xi d t\\
	&+\sum_{i=1}^{n} \sum_{j=1}^{n}\int_{0}^{1} \int_{0}^{1}(1-\xi)\mathrm{E}\left[\partial_{i}\partial_{j} m\left(\tilde H^{(ij)}(t)+\xi e_i^{(ij)}(t)+\xi e_j^{(ij)}(t)\right) \dot{H}_{i j}(t)h_j^{(ij)}(t)\right] d \xi d t\\
	&\doteq \Delta_{11}+\Delta_{12}+\Delta_{13}.
\end{align*}
Note that the random vector $\tilde H^{(ij)}(t)$ is independent of $\big(\dot{H}_{i j}(t), H_{i j}(t)\big)$ by its construction, and $\mathrm{E}\left[\dot{H}_{i j}(t)\right]=0$.
Hence, we have $\Delta_{11}=0$.


By the construction of $\tilde{H}^{(ij)}(t)$, it is independent of $\dot{H}_{ij}(t)$ and $h_k^{(ij)}(t)$. Moreover, by the definition of $\dot{H}_{ij}(t)$ and $h_k^{(ij)}(t) \ (k \in \{i,j\})$, we have $E[\dot{H}_{ij}(t)h_i^{(ij)}(t)]=n^{-1}\sum_{l \neq i,j} E[
V_{ij}V_{il}-y_{ij}y_{il}]=n^{-1}\sum_{l \neq i,j} E(
V_{ij}V_{il})$. If $j \notin \mathcal{C}_{g_i}$, we have
\[
E(V_{ij}V_{il})= \left\{
\begin{aligned}
	&\frac{n}{(\frac{1}{n-n_{g_i}}+\frac{1}{n_{g_i}-1})(p_{22}-p_{12})} \cdot \frac{p_{12}}{(n-{n_{g_i}})^2}, & \quad  \text{if} \ l \notin \mathcal{C}_{g_i}, \\
	&\frac{n}{(\frac{1}{n-n_{g_i}}+\frac{1}{n_{g_i}-1})(p_{22}-p_{12})} \cdot \frac{-p_{12}}{(n-{n_{g_i}})(n_{g_i}-1)}, & \quad  \text{if} \ l \in \mathcal{C}_{g_i},
\end{aligned}
\right.
\]
This implies that
\[
\begin{aligned}
	n^{-1}\sum_{k \neq i,j} E(
	V_{ij}V_{il})&=\frac{p_{12}}{(\frac{1}{n-n_{g_i}}+\frac{1}{n_{g_i}-1})(p_{22}-p_{12})(n-{n_{g_i}})}
	\Big(\frac{n-n_{g_i}-1}{n-n_{g_i}}-\frac{n_{g_i}-1}{n_{g_i}-1}\Big) \\
	&=\frac{-p_{12}}{(\frac{1}{n-n_{g_i}}+\frac{1}{n_{g_i}-1})(p_{22}-p_{12})(n-{n_{g_i}})^2} =O(n^{-1}).
\end{aligned}
\]
If $j \in \mathcal{C}_{g_i}$, it can be easily proved that $n^{-1}\sum_{k \neq i,j} E(
V_{ij}V_{il})=O(n^{-1})$. Therefore, we have
\[
\begin{aligned}
	|\Delta_{12}|&=\Bigg|\sum_{i=1}^{n}\sum_{j=1}^{n}
	\sum_{k \in \{i,j\}}
	\int_{0}^{1} E\Big[ \partial_i\partial_k m\big(\tilde{H}^{(ij)}(t)\big)
	\Big]E\big[\dot{H}_{ij}(t)h_k^{(ij)}(t)\big]dt\Bigg|  \\
	&\lesssim n^{-1} \Bigg|\sum_{i=1}^{n}\sum_{j=1}^{n}
	\sum_{k \in \{i,j\}}
	\int_{0}^{1} E\Big[ \partial_i\partial_k m\big(\tilde{H}^{(ij)}(t)\big)
	\Big]dt\Bigg|  \\
	& \lesssim n^{-1} \sum_{i=1}^{n}\sum_{j=1}^{n} \int_{0}^{1} E[U_{ij}(\tilde{H}^{(ij)}(t))] dt \\
	& \le n^{-1} (G_2+2G_1 \kappa),
\end{aligned}
\]
where the first equality is by the independence of $\tilde{H}^{(ij)}(t)$ and $\dot{H}_{ij}(t)h_k^{(ij)}(t)$, the third and forth inequalities are by Lemma A.5 in Chernozhukov et al. (2013). Using the fact that $G_1 \lesssim \psi$, $G_2 \lesssim \psi^2$, $\psi=n^{1/8}$ and $\kappa=\psi \log n$, we obtain that $\Delta_{12}=O(\psi^2\log n/n)=o(1)$.

Observe that
\[
\begin{aligned}
	|\Delta_{13}| &\le \sum_{i=1}^{n}\sum_{j=1}^{n}
	\sum_{k,l \in \{i,j\}}
	\int_{0}^{1}
	E\Big[
	U_{ikl}(\tilde{H}^{(ij)}(t))\cdot |  \dot{H}_{ij}(t)h_k^{(ij)}(t)h_l^{(ij)}(t)|
	\Big] dt \\
	& =\sum_{i=1}^{n}\sum_{j=1}^{n}
	\sum_{k,l \in \{i,j\}}
	\int_{0}^{1}
	E\Big[
	U_{ikl}(\tilde{H}^{(ij)}(t))\Big] \cdot E\Big[ |  \dot{H}_{ij}(t)h_k^{(ij)}(t)h_l^{(ij)}(t)|
	\Big]dt  \\
	& \lesssim n^{-1} (G_3+G_2\kappa+G_1 \kappa^2)
	\int_{0}^{1}
	E\Big[ \max_{1\le i, j \le n, k,l \in \{i,j\}} |  \dot{H}_{ij}(t)h_k^{(ij)}(t)h_l^{(ij)}(t)|
	\Big]dt  \\
	& \le n (G_3+G_2\kappa+G_1 \kappa^2)
	\int_{0}^{1}
	E\Big[ \max_{1\le i, j, k, l, s_1, s_2 \le n,} |  \dot{H}_{ij}(t)
	H_{ks_1}^{(ij)}(t)H_{ls_2}^{(ij)}(t)|
	\Big]dt \\
	& \le n^{-1/2} (G_3+G_2\kappa+G_1 \kappa^2) \bar{E} \big[ \max_{ 1 \le j \le n}
	(|V_{ij}|+|y_{ij}|)^3
	\big],
\end{aligned}
\]
where the first inequality is by Lemma A.5 and Lemma A.6 in Chernozhukov et al. (2013), the second equality is by the independence of $U_{ikl}(\tilde{H}^{(ij)}(t))$ and $\dot{H}_{ij}(t)h_k^{(ij)}(t)h_l^{(ij)}(t)$, the third inequality is due to the fact that $\sum_{i=1}^{n}\sum_{j=1}^{n}U_{ijk}(z)=O\{n^{-1}(G_3+G_2\kappa+G_1 \kappa^2)\}$, the forth equality is by the definition of $h_k^{(ij)}(t)$, and the last equality is by Holder inequality. Combining above results, we have that
\[
|\Delta_1| \le n^{-1/2} (G_3+G_2\kappa+G_1 \kappa^2) \bar{E} \big[ \max_{ 1 \le j \le n}
(|V_{ij}|+|y_{ij}|)^3\big]+o(1).
\]
Note that $V_{ij,\tau}$ is bounded and $y_{ij}\sim_{iid} N(0,1)$. These, together with the Bonferroni inequality, imply that
$\overline{\mathrm{E}}\left[\max _{1 \leqslant j \leqslant n}\left(\left|V_{i j,\tau}\right|+\left|y_{i j}\right|\right)^{2}\right]=O(\log n)$.
Using the results that
$G_{1}\lesssim \psi$ and $G_{2} \lesssim \psi^{2}$ with $\psi=n^{1/8}$ and $\kappa=\psi \log n$,
we further  obtain $\Delta_1=o(1)$.

We next consider $\Delta_2$, which can be expressed as
\begin{align*}
	\Delta_2 &=\sum_{i, k=1}^{n} \sum_{j=1}^{n} \int_{0}^{1} \mathrm{E}\left[\partial_{i} \partial_{k} m\left(H^{(ij)}(t)\right) \dot{H}_{i j}(t) \{H_{kj }(t)+H_{ki}(t)\}\right] d t \\
	&=\sum_{i, k=1}^{n} \sum_{j=1}^{n} \int_{0}^{1} \mathrm{E}\left[\partial_{i} \partial_{k} m\left(H^{(ij)}(t)\right) \dot{H}_{i j}(t) H_{ki}(t)\right] d t\\
	&+\sum_{i, k=1}^{n} \sum_{j=1}^{n} \int_{0}^{1} \mathrm{E}\left[\partial_{i} \partial_{k} m\left(H^{(ij)}(t)\right) \dot{H}_{i j}(t) H_{kj }(t)\right] d t\doteq \Delta_{21}+\Delta_{22}.
\end{align*}
Here, we only focus on $\Delta_{21}$ since the proof of $\Delta_{22}$ is similar.
Define $H^{(ijk)}(t)=H^{(ij)}(t)-H_k(t)\doteq (h_1^{(ijk)(t)},\cdots,h_n^{(ijk)(t)})$ and $\tilde H^{(ijk)}(t)=H^{(ijk)}(t)-e_i^{(ijk)}-e_j^{(ijk)}-e_k^{(ijk)}$,
where $e_l^{(ijk)}=(0,\cdots,0,H^{(ijk)}_l(t),0,\cdots,0)^\top$. Employing Taylor's theorem, we obtain that
\begin{align*}
	\Delta_{21}=&\sum_{i, k=1}^{n} \sum_{j=1}^{n} \int_{0}^{1} \mathrm{E}\left[\partial_{i} \partial_{k} m\left(H^{(ijk)}(t)\right) \dot{H}_{i j}(t) H_{kj }(t)\right] d t\\
	=&\sum_{i, k=1}^{n} \sum_{j=1}^{n} \int_{0}^{1} \mathrm{E}\left[\partial_{i} \partial_{k} m\left(\tilde H^{(ijk)}(t)\right) \dot{H}_{i j}(t) H_{kj }(t)\right] d t \\
	+&\sum_{i, k=1}^{n} \sum_{j=1}^{n} \int_{0}^{1} \int_{0}^{1} (1-\xi)\mathrm{E}\Bigg[\partial_{i} \partial_{k}\partial_{i} m\left(\tilde H^{(ijk)}(t)+\xi e_i^{(ijk)}(t)+\xi e_j^{(ijk)}(t)+\xi e_k^{(ijk)}(t)\right)\\
	\times & \dot{H}_{i j}(t) H_{kj }(t)e_i^{(ijk)}\Bigg] d\xi d t \\
	+&\sum_{i, k=1}^{n} \sum_{j=1}^{n} \int_{0}^{1} \int_{0}^{1} (1-\xi)\mathrm{E}\Bigg[\partial_{i} \partial_{k}\partial_{j} m\left(\tilde H^{(ijk)}(t)+\xi e_i^{(ijk)}(t)+\xi e_j^{(ijk)}(t)+\xi e_k^{(ijk)}(t)\right)\\
	\times & \dot{H}_{i j}(t) H_{kj }(t)e_j^{(ijk)}\Bigg] d\xi d t\\
	+&\sum_{i, k=1}^{n} \sum_{j=1}^{n} \int_{0}^{1} \int_{0}^{1} (1-\xi)\mathrm{E}\Bigg[\partial_{i} \partial_{k}\partial_{k} m\left(\tilde H^{(ijk)}(t)+\xi e_i^{(ijk)}(t)+\xi e_j^{(ijk)}(t)+\xi e_k^{(ijk)}(t)\right)\\
	\times&\dot{H}_{i j}(t) H_{kj }(t)e_k^{(ijk)}\Bigg] d\xi d t\\
	\doteq & \Delta_{21,1}+\Delta_{21,2}+\Delta_{21,3}+\Delta_{21,4}.
\end{align*}

We evaluate the above four terms separately.
By the construction of $\left(Y_{i}\right)_{i=1}^{n}$,
we have $\mathrm{E}\left[\dot{H}_{i j}(t) H_{kj}(t)\right]=$
$n^{-1} \mathrm{E}\left[V_{i j,\tau} V_{kj,\tau}-y_{i j} y_{kj}\right]=0$.
In addition,  by the definition of
$\tilde H^{(ijk)}$, it is independent of $H_{kj}(t)$ and $\dot{H}_{ij}(t)$. The above results imply that  $\Delta_{21,1}=0$.
By Lemma A.5 in Chernozhukov et al. (2013), we have
\begin{align*}
	\Delta_{21,3} &\lesssim (G_3+G_2\kappa+G_1\kappa^2)n\int \overline{\mathrm{E}}\Big\{\max_{1\le i,j,k\le n}\dot{H}_{i j}(t) H_{kj }(t)e_j^{(ijk)}\big\} dt\\
	&\lesssim  n^{-1/2}(G_3+G_2\kappa+G_1\kappa^2)\overline{\mathrm{E}}\left[\max _{1 \leqslant j \leqslant n}\left(\left|V_{i j,\tau}\right|+\left|y_{i j}\right|\right)^{3}\right].
\end{align*}
Analogous results can be obtained for $\Delta_{21,2}$ and $\Delta_{21,4}$.
These, together with the above result of $\Delta_{21,1}$, immediately imply that there exists a finite positive constant
$C_3^{(1)}$ such that
\beq
\Delta_2\le C_3^{(1)}\left(\psi^{3}+\kappa \psi^{2}+\kappa^{2} \psi\right)\left(n^{-1 / 2} \overline{\mathrm{E}}\left[S_{i}^{3}\right]\right). \label{A.6}
\eeq

We lastly evaluate $\Delta_3$ by showing that
\begin{equation}
	|\Delta_3|\lesssim\left(G_{3}+G_{2} \kappa+G_{1} \kappa^{2}\right) n \int \overline{\mathrm{E}}\left[\max _{1 \leqslant j, k, l \leqslant n}\left|\dot{H}_{i j}(t) H_{i k}(t) H_{i l}(t)\right|\right] d t \label{A.7}
\end{equation}
\begin{equation}
	\lesssim  n^{-1 / 2}\left(G_{3}+G_{2} \kappa+G_{1} \kappa^{2}\right) \overline{\mathrm{E}}\left[\max _{1 \leqslant j \leqslant n}\left(\left|V_{i j,\tau}\right|+\left|y_{i j}\right|\right)^{3}\right]. \label{A.8}
\end{equation}

Using Lemma A.5 in  Chernozhukov et al. (2013), we have
that $\left|\partial_{j} \partial_{k} \partial_{l} m\left(z\right)\right| \lesssim$
$\left(G_{3}+G_{2} \kappa+G_{1} \kappa^{2}\right)$. Hence (\ref{A.7}) holds.
To show (\ref{A.8}),
define $\omega(t)=1 /\min\{\sqrt{t}, \sqrt{1-t}\}$. We then have
\begin{align*}
	&\int_{0}^{1} n \overline{\mathrm{E}}\left[\max _{1 \leqslant j, k, l \leqslant p}\left|\dot{H}_{i j}(t) H_{i k}(t) H_{i l}(t)\right|\right] d t \\
	&=\int_{0}^{1} \omega(t) n \overline{\mathrm{E}}\Big[\max _{1 \leqslant j, k, l \leqslant p} \mid \big(\dot{H}_{i j}(t) / \omega(t)\big) H_{i k}(t) H_{i l}(t) \mid\Big] d t\\
	&\leqslant n \int_{0}^{1} \omega(t)\left(\overline{\mathrm{E}}\left[\max _{1 \leqslant j \leqslant p}\left|\dot{H}_{i j}(t) / \omega(t)\right|^{3}\right] \overline{\mathrm{E}}\left[\max _{1 \leqslant j \leqslant p}\left|H_{i j}(t)\right|^{3}\right] \overline{\mathrm{E}}\left[\max _{1 \leqslant j \leqslant p}\left|H_{i j}(t)\right|^{3}\right]\right)^{1 / 3} d t \\
	&\leqslant n^{-1 / 2}\left\{\int_{0}^{1} \omega(t) d t\right\} \overline{\mathrm{E}}\left[\max _{1 \leqslant j \leqslant p}\left(\left|V_{i j,\tau}\right|+\left|y_{i j}\right|\right)^{3}\right],
\end{align*}
where the first inequality follows  Holder's inequality, and the second inequality
is due to the fact that
$\left|\dot{H}_{i j}(t) / \omega(t)\right| \leqslant\left(\left|V_{i j,\tau}\right|+\left|y_{i j}\right|\right) / \sqrt{n},\left|H_{i j}(t)\right| \leqslant\left(\left|V_{i j,\tau}\right|+\right.$
$\left.\left|y_{i j}\right|\right) / \sqrt{n}$.
This, together with $\int_{0}^{1} \omega(t) d t \lesssim 1$, leads to (\ref{A.8}).

By (\ref{A.6})--(\ref{A.8}) and $\Delta_1=o(1)$, we have demonstrated (\ref{A.5}).
Note that $V_{ij,\tau}$ is bounded and $y_{ij}\sim_{iid} N(0,1)$. These, together with the Bonferroni inequality, implies that
$$\overline{\mathrm{E}}\left[\max _{1 \leqslant j \leqslant n}\left(\left|V_{i j,\tau}\right|+\left|y_{i j}\right|\right)^{3}\right]=O(\log^{3/2} n).$$
By (\ref{A.2})--(\ref{A.5}), we obtain that
$$
\begin{aligned}
	\mathrm{P}\left(T_{0} \leqslant t\right) & \leqslant \mathrm{P}\left(F_{\kappa}(Q) \leqslant t+e_{\kappa}\right) \leqslant \mathrm{E}\big\{g\left(F_{\kappa}(Q)\right)\big\} \\
	& \leqslant \mathrm{E}\left[g\left(F_{\kappa}(Y)\right)\right]+C_3\left(\psi^{3}+\kappa \psi^{2}+\kappa^{2} \psi\right)\left(n^{-1 / 2} \overline{\mathrm{E}}\left[S_{i}^{3}\right]\right)+o(1) \\
	& \leqslant \mathrm{P}\left(F_{\kappa}(Y) \leqslant t+e_{\kappa}+\psi^{-1}\right)+C_3\left(\psi^{3}+\kappa \psi^{2}+\kappa^{2} \psi\right)\left(n^{-1 / 2} \overline{\mathrm{E}}\left[S_{i}^{3}\right]\right)+o(1)\\
	& \leqslant \mathrm{P}\left(Z_{0} \leqslant t+e_{\kappa}+\psi^{-1}\right)+C_3\left(\psi^{3}+\kappa \psi^{2}+\kappa^{2} \psi\right)\left(n^{-1 / 2} \overline{\mathrm{E}}\left[S_{i}^{3}\right]\right)+o(1) \\
	&\le P(Z_0\le t)+C_3(e_{\kappa}+\psi^{-1})\sqrt{\log(n\psi)}+C_2\left(\psi^{3}+\kappa \psi^{2}+\kappa^{2} \psi\right)\left(n^{-1 / 2} \overline{\mathrm{E}}\left[S_{i}^{3}\right]\right)+o(1)\\
	&=P(Z_0\le t)+O(n^{-1/8}\log n)+O(n^{-1/8}\log^{5/2}n)+o(1)=P(Z_0\le t)+o(1).
\end{aligned}
$$
Consequently, we have shown that, for any $t\in \mathbb{R}$,
\beq
P\big(\max_{1\le i\le n}Q_{i,\tau} \leq t\big)-P\big(Z_0\le t\big)\to 0, \label{A.9}
\eeq
which proves (\ref{A.1}) and thus complete the proof of Step I.

{\sc Step II}.
Under the null hypothesis of $H_0$, we employ the
Bonferroni and Bernstein inequalities and obtain that $\max_{1\le i\le n}|\hat p_0^{(i)}-p_0|=O_p(\sqrt{\log n/{n}})$
and $\max_{1\le i\le n}|\hat p_{12}^{(i)}-p_{12}|=O_p(\sqrt{\log n/{n}})$.
In addition, by (\ref{A.9}), we have $\max_{1\le i\le n}|Q_{i,\tau}|=O_p(\log^{1/2} n)$.
As a result, we obtain that there exist a finite positive constant $C_4$ such that
\begin{align*}
	&|\max_{1\le i\le n}Q_{i,\tau} -\max_{1\le i\le n}T_{i,\tau}|\nonumber\\ \nonumber
	&=\left|\max_{1\le i\le n}\frac{\hat{p}_i^{bet}-\hat{p}^{in}_i}{\{(\frac{1}{n-n_{g_i}}+\frac{1}{n_{g_i}-1})(p_{22}-p_{12})\}^{1/2}}-\max_{1\le i\le n}\frac{\hat{p}_i^{bet}-\hat{p}^{in}_i}{\{(\frac{1}{n-n_{g_i}}+\frac{1}{n_{g_i}-1})(\hat{p}^{(i)}_{22}-\hat{p}^{(i)}_{12})\}^{1/2}}\right|\\
	&\le C_4 \max_{1\le i\le n} |Q_{i,\tau}| \max_{1\le i\le n}\big|\{p_{22}-p_{12}\}^{-1/2}-(\hat{p}^{(i)}_{22}-\hat{p}^{(i)}_{12})^{-1/2}\big|=O_p\Big(\frac{\log n}{n^{1/2}}\Big)\to 0.
\end{align*}
This immediately leads to, for any
$t\in \mathbb{R}$,
\begin{align}
	P\big(\max_{1\le i\le n}Q_{i,\tau} \leq t\big)-P\big(\max_{1\le i\le n}T_{i,\tau}  \le t\big)\to 0. \label{A.10}
\end{align}

Under the null hypothesis of $H_0$, we can also employ the Bernstein inequality to
obtain  that $\max_{1\le i_1,i_2\le n}|\hat\sigma_{i_1i_2}-\sigma_{i_1i_2}|=O_p(n^{-1/2})$.
Subsequently, we employ Lemma 1 by setting $\Sigma^V=\hat\Sigma$, $\Sigma^U=\Sigma$,
$\Delta_0=O_p(n^{-1/2})$, $C_1=2\lambda_{\max}(\Sigma)$ and $c_1=\lambda_{\min}(\Sigma)/2$,
and then have that, for any $t\in \mathbb{R}$,
\begin{align}
	P\big(Z_0\le t\big)-P\big(\tilde Z_0\le t\big) \to 0, \label{A.11}
\end{align}
where $\tilde Z_0=\max_{1\le i\le n} Z_i$ and $Z=(Z_1,\cdots, Z_n)^\top$ is a multivariate normally
distributed random variable with mean zero and covariance matrix $\Sigma$.
By (\ref{A.9})--(\ref{A.11}), we obtain
that
\begin{align}
	P\big(\max_{1\le i\le n}T_{i,\tau}  \le t\big)-P\big(\tilde Z_0\le t\big) \to 0. \label{A.12}
\end{align}
By taking the same procedure as in Chernozhukov et al. (2017),
we can prove that the Gaussian approximation could hold in all hyper-rectangles of $\mathbb{R}^n$. Thus, by considering the special hyper-rectangles $[-t,t]^n$, we have
\begin{align}
	P\big(\max_{1\le i\le n}|T_{i,\tau}|  \le t\big)-P\big(\max_{1\le i\le n} |Z_i|\le t\big) \to 0, \label{A.122}
\end{align}
which completes the entire proof.

\subsection*{S.2.2 Proof of Theorem 2}

We prove this theorem in two steps. The first step demonstrates that
$z_{\alpha}<(2+\kappa/2)\log n$ with probability tending to one for some finite constant $\kappa>0$, where $z_{\alpha}$ was defined in Theorem 2.
The second step shows that $P\big\{T_{\tau}>(2+\kappa/2)\log n\big\}\rightarrow 1$.

{\sc STEP I.} Recall that $O_b^{(l)}=(O_{b1}^{(l)}, \cdots, O_{bn}^{(l)})^\top$, for
$l=1,\cdots, N$, are independently generated from the multivariate normal distribution with mean zero and covariance
matrix $\hat\Sigma$.  Then, we have $O_{bi}^{(l)}$s are standard normal variables, for $i=1,\cdots, n$, since the test statistics $T_{i,\tau}$ are  standardized and the diagonal
elements of $\hat\Sigma$ are 1. Subsequently,
by the Bonferroni inequality and the exponential tail probability of the standard normal distribution,  we obtain that
\begin{align*}
	P\big\{\max_i {O_{bi}^{(l)}}^2>(2+\kappa/2)\log n\big\}\le nP\big\{ {O_{bi}^{(l)}}^2>(2+\kappa/2)\log n\big\}\to 0,
\end{align*}
which implies that $z_{\alpha}<(2+\kappa/2)\log n$ with probability tending to one, and completes the first step of the proof.

{\sc STEP II.} We next show that $P\big\{T_{\tau}>(2+\kappa/2)\log n\big\}\to 1$.
Define
\begin{align*}
	\tilde Q_{i,\tau}=&\frac{\hat{p}_{i}^{bet}-\hat{p}^{in}_{i}-\left(\sum_{k\not=g_i}\frac{\gamma_{k}}{1-\gamma_{g_i}}p_{g_ik,\tau}-p_{g_ig_i,\tau}\right)}{\big\{(\frac{1}{n-n_{g_i}}+\frac{1}{n_{g_i}-1})((\sum_{k=1}^K \gamma_k p_{g_ik,\tau})(1-\sum_{k=1}^K \gamma_k p_{g_ik,\tau})-\tilde{\delta}_{12i})\big\}^{1/2}}, \mbox{~and~}\\
	\breve Q_{i,\tau}=&\frac{\hat{p}_{i}^{bet}-\hat{p}^{in}_{i}}{\big\{(\frac{1}{n-n_{g_i}}+\frac{1}{n_{g_i}-1})((\sum_{k=1}^K \gamma_k p_{g_ik,\tau})(1-\sum_{k=1}^K \gamma_k p_{g_ik,\tau})-\tilde{\delta}_{12i})\big\}^{1/2}}.
\end{align*}
Employing the same procedure as that used in the proof of Theorem 1, we  obtain that
\begin{align*}
	P\Big(\max_{1\le i\le n} \tilde{Q}_{i,\tau}<x\Big)-P\Big(\max_{1\le i\le n} J_{i}<x\Big)\to 0
\end{align*}
for any $x\in \mathbb{R}$, where $(J_1,\cdots,J_n)^\top$ follows the multivariate normal distribution with mean zero and the same covariance matrix as that of $(Q_{1,\tau},\cdots,Q_{n,\tau})^\top$.
Thus, we have $$P\Big(\max_{1\le i\le n}\tilde Q_{i,\tau}^2>(2+\kappa/2)\log n\Big)\to 0.$$
Under the alternative hypothesis of $H_1$, we employ the
Bonferroni and Bernstein inequalities and obtain that $\max_{1\le i\le n}|\hat p_{0}^{(i)}-\sum_{k=1}^K \gamma_k p_{g_ik,\tau}|=O_p(\sqrt{\log n/{n}})$
and $\max_{1\le i\le n}|\hat p_{12}^{(i)}-\tilde{\delta}_{12i}|=O_p(\sqrt{\log n/{n}})$.
Subsequently, by the triangle inequality, we have
\begin{align*}
	&P\Big(\max_{1\le i\le n}T_{i,\tau}^2>(2+\kappa/2)\log n\Big)\ge P\Bigg\{\Big(1-O\big(\sqrt{\frac{\log n}{n}}\big)\Big)\max_{1\le i\le n}\breve Q_{i,\tau}^2>(2+\kappa/2)\log n\Bigg\}\\
	&\ge P\Big(\frac{1}{2}\max_{1\le i\le n}n\nu_i-\max_{1\le i\le n}\tilde Q_{i,\tau}^2>(2+\kappa/2)\log n\Big)\to 1,
\end{align*}
where the last inequality is due to the theorem condition, $\max_{1\le i\le n}n\nu_i>(4+\kappa)\log n$. This completes the entire proof.

\subsection*{S.2.3 Proof of Theorem 3}

Define $\breve Z=(\breve Z_1,\cdots, \breve Z_{n})={\hat \Sigma}^{-1/2}(\breve Y_1,\cdots, \breve Y_{n})$, where
$\breve Y_i=\frac{1}{\sqrt{n}}\sum_{j=1}^{n} \breve y_{ij}$, $\breve y_{ij}$s
are independent and identically distributed standard normal variables, and the covariance matrix of
$\breve Y\doteq(\breve Y_1,\cdots, \breve Y_{n})$ is $\Sigma$.
We next prove the theorem in two steps. In the first step, we show that
for any $x\in\mR$,
\begin{align}
	P\big(\max_{1\leq i\leq n} \breve Z_{i}\le x\big)-P\big(\max_{1\le i\le n} \mM^{adj}_{i,\tau}\le x\big)\to 0. \label{A.13}
\end{align}
In the second step, we demonstrate that the difference between $\hat \Sigma$ and $\Sigma$ is negligible.

To prove (\ref{A.13}), it is equivalent to show that
\begin{align}
	P\big(\breve Y\le x{\hat \Sigma}^{1/2}\bm 1\big)-P\big(Q\le x{\hat \Sigma}^{1/2}\bm 1 \big)\to 0,  \label{A.14}
\end{align}
where $Q$ was defined in the proof of Theorem 1.
To verify (\ref{A.14}), it suffices to show
\begin{align}
	P\big(||\breve Y+\delta||_{\infty}\le x\big)-P\big(||Q+\delta||_{\infty}\le x\big)\to 0 \label{A.15}
\end{align}
for $\delta=(\delta_1,\cdots,\delta_{n}) =x(I_n-{\hat \Sigma}^{1/2})\bm 1$.
Analogous to the proof of Theorem 1, we only need to prove that $
E\big[m(Q+\delta)-m(\breve Y+\delta)\big]\to 0$, where $m(\cdot)$ was defined
in the proof of Theorem 1.

To verify the above result, we
define $W_k=(W_{k1},\cdots,W_{kn})^\top$, where $W_{ki}=\frac{1}{\sqrt{n}}\sum_{j=1}^k V_{ij,\tau}+\frac{1}{\sqrt{n}}\sum_{j=k+1}^{n} \breve y_{ij}+\delta_i$ and $  V_{ij,\tau}$ was defined in the proof of Theorem 1.
Accordingly, we have $W_{n}=Q+\delta$ and $W_0=\breve Y+\delta$. Thus,
\begin{align*}
	E[m(Q+\delta)-m(\breve Y+\delta)]&=\sum_{k=1}^{n}E[m(W_k)-m(W_{k-1})]\\
	&=\sum_{k=1}^{n}\Big\{E[m(W_k)-m(W_{0k})]-E[m(W_{k-1})-m(W_{0k})]\Big\},
\end{align*}
where $W_{0k}=(W_{0k1},\cdots,W_{0kn})^\top$
with $W_{0ki}=\frac{1}{\sqrt{n}}\sum_{j=1}^{k-1} V_{ij,\tau}+\frac{1}{\sqrt{n}}\sum_{j=k+1}^{n} \breve y_{ij}+\delta_i$.
Employing Taylor's theorem, we obtain that
\begin{align}
	&E[m(W_k)-m(W_{0k})]\n\\
	&=\sum_{i=1}^{n}\partial_i m(W_{0k})^\top (W_{ki}-W_{0ki})+\frac{1}{2}\sum_{i=1}^{n}\sum_{l=1}^{n}\partial_i\partial_l m(W_{0k})^\top (W_{ki}-W_{0ki})(W_{kl}-W_{0kl})\n\\
	&+\frac{1}{6}\sum_{i=1}^{n}\sum_{l=1}^{n}\sum_{s=1}^{n}\partial_i\partial_l\partial_s m(W_{0k}+\breve\xi_1(W_k-W_{0k}))^\top (W_{ki}-W_{0ki})(W_{kl}-W_{0kl})(W_{ks}-W_{0ks})\n\\
	&=\frac{1}{\sqrt{n}}\sum_{i=1}^{n}\partial_i m(W_{0k})^\top   V_{ik,\tau}+\frac{1}{2n}\sum_{i=1}^{n}\sum_{l=1}^{n}\partial_i\partial_l m(W_{0k})^\top   V_{ik,\tau}  V_{il,\tau}\n\\
	&+\frac{1}{6n^{3/2}}\sum_{i=1}^{n}\sum_{l=1}^{n}\sum_{s=1}^{n}\partial_i\partial_l\partial_s m(W_{0k}+\breve\xi_1(W_k-W_{0k}))^\top   V_{ik,\tau}  V_{il,\tau}  V_{is,\tau}, \label{A.16}
\end{align}
where $\breve\xi_1$ is a constant within $(0,1)$. Analogously, for some $\breve\xi_2\in (0,1)$, we obtain that
\begin{align}
	&E[m(W_{k-1})-m(W_{0k})]\n \\
	&=\frac{1}{\sqrt{n}}\sum_{i=1}^{n}\partial_i m(W_{0k})^\top \breve y_{ik}+\frac{1}{2n}\sum_{i=1}^{n}\sum_{l=1}^{n}\partial_i\partial_l m(W_{0k})^\top   \breve y_{ik}  \breve y_{il}\n \\
	&+\frac{1}{6n^{3/2}}\sum_{i=1}^{n}\sum_{l=1}^{n}\sum_{s=1}^{n}\partial_i\partial_l\partial_s m(W_{0k}+\breve\xi_2(W_k-W_{0k}))^\top \breve y_{ik} \breve y_{il} \breve y_{is}. \label{A.17}
\end{align}
By (\ref{A.16}) and (\ref{A.17}), we then have
\begin{align*}
	&E[m(W_k)-m(W_{0k})]-E[m(W_{k-1})-m(W_{0k})]\\
	=&E\left\{\frac{1}{\sqrt{n}}\sum_{i=1}^{n}\partial_i m(W_{0k})^\top (  V_{ik,\tau}- \breve y_{ik})\right\}\\
	&+E\left\{\frac{1}{2n}\sum_{i=1}^{n}\sum_{l=1}^{n}\partial_i\partial_l m(W_{0k})^\top (  V_{ik,\tau}  V_{il,\tau}- \breve y_{ik} \breve y_{il})\right\}\\
	&+E\left\{\frac{1}{6n^{3/2}}\sum_{i=1}^{n}\sum_{l=1}^{n}\sum_{s=1}^{n}\partial_i\partial_l\partial_s m(W_{0k}+\breve\xi_1(W_k-W_{0k}))^\top V_{ik,\tau}  V_{il,\tau}  V_{is,\tau}\right\}\\
	&-E\left\{\frac{1}{6n^{3/2}}\sum_{i=1}^{n}\sum_{l=1}^{n}\sum_{s=1}^{n}\partial_i\partial_l\partial_s m(W_{0k}+\breve\xi_2(W_k-W_{0k}))^\top \breve y_{ik}  \breve y_{il} \breve y_{is}\right\}\\
	\doteq &\widetilde\Delta_1+\widetilde{\Delta}_2+\widetilde{\Delta}_3+\widetilde{\Delta}_4.
\end{align*}
Employing the similar techniques to those used in the proofs of
$\Delta_1,\Delta_2$ and $\Delta_3$ in  Theorem 1, we can show that $\widetilde{\Delta}_s=O(n^{-3/2}\zeta^2\log^{3/2} n)$ for any $s=1,2,3,4$ with
$\zeta=n^{1/8} \log n$.
Using these results, we then have $$E[m(Q+\delta)-m(\breve Y+\delta)]=\sum_{k=1}^{n}E\big\{m(W_k)-m(W_{k-1})\big\}=O(n^{-1/2}\zeta^2\log^{3/2} n)=o(1),$$
which completes the proof of the first step.

In the second step, we need to
evaluate the difference between $\hat \Sigma$ and $\Sigma$.
For any matrix $A=(a_{ij})\in \mathbb{R}^{p\times p}$,
denote $||A||_m=\max_{1\le i,j\le p}|a_{ij}|$.
By the condition of Theorem 3, the eigenvalues of $\hat \Sigma$ and $\Sigma$ are all bounded from 0 to infinity. Then, by the Bonferroni inequality, we have
\begin{align*}
	\big\|\hat \Sigma^{-1/2}\Sigma^{1/2}-I_{n}\big\|_m=\big\|\hat \Sigma^{1/2}(\Sigma^{1/2}+\hat \Sigma^{1/2})^{-1}(\Sigma- \hat \Sigma)\big\|_m
	=O_p\big(\big\|\hat \Sigma-\Sigma\big\|_m\big)
\end{align*}
because $\lambda_{max}(\hat \Sigma^{1/2}(\Sigma^{1/2}+\hat \Sigma^{1/2})^{-1})\le 1$.
Further applying the Bernstein inequality, we have $\max_{i_1,i_2}|\hat{\sigma}_{i_1i_2}-\sigma_{i_1i_2}|=O_p(n^{-1/2})$,
where $\hat{\sigma}_{i_1i_2}$ and $\sigma_{i_1i_2}$ are elements of $\hat \Sigma$ and $\Sigma$, respectively.
Accordingly, we have that $\big\|\hat \Sigma^{-1/2}\Sigma^{1/2}-I_{n}\big\|_m=O_p\big\{n^{-1/2}\big\}$, which completes the proof of the second step.

Based on the above result, we can set  $\Delta_0=O_p(n^{-1/2})$ in Lemma 1. In addition, set $\Sigma^V=\hat \Sigma^{-1/2}\Sigma^{1/2}$,
$\Sigma^U=I_{n}$, $C_1=1$ and $c_1=0.5$.
Employing  Lemma 1, we then have, for any $t\in\mR^{+}$,
\begin{align}
	P(\max_{1\leq i\leq n} \breve Z_{i}^2\le t)-P(\max_{1\leq i\leq n} Z_{Ii}^2\le t)\to 0, \label{A.18}
\end{align}
where $Z_I=(Z_{I1},\cdots, Z_{In})^\top$ is a multivariate normally distributed random variable with mean zero  and covariance matrix $I_n$.
This, in conjunction with (\ref{A.13}), leads to
\begin{align}
	P\left(\max_{1\leq i\leq n} Z_{Ii}^2\le t\right)-P\left( T^{adj}_{\tau}\le t\right)\to 0. \label{A.19}
\end{align}
In addition, by Lemma 6 of Cai et al. (2014), we have that, for any $x\in\mR$,
\begin{align*}
	P\Big\{\max_{1\leq i\leq n} Z_{Ii}^2-2\log(n)+\log\log(n)\le x\Big\}\to \exp\Big\{-\frac{1}{\sqrt{\pi}}\exp(-x/2)\Big\}.
\end{align*}
This, together with  (\ref{A.19}), completes the entire proof.

\subsection*{S.2.4 Proof of Theorem 4}

The proof of Theorem 4 is similar to the proof of Theorem 2, except the signals $\nu_i$ are replaced by $\omega_i^2$. The details are omitted.

\subsection*{S.2.5 Proof of Theorem 5}
We only show  part (i) of this theorem, since the proof of part (ii) is similar.
Let \begin{align*}
	\tilde T^{\epsilon}_{\tau}=\max_{1\le i\le n} \frac{(\hat{p}_{i}^{\hat\epsilon,bet}-\hat{p}^{\hat\epsilon, in}_{i})^2}{(\frac{1}{n-n_{g_i}}+\frac{1}{n_{g_i}-1})(p^{\epsilon}_{22}-p^{\epsilon}_{12})}.
\end{align*}
Employing the same techniques as those used in the proof of (10) in the main paper,  we only need to show that,  $x\in\mR^+$,
\begin{align}\label{A.20}
	P\big(\tilde T^{\epsilon}_{\tau}<x\big)-P(\max_{1\le i\leq n} (Z_i^{\epsilon})^2<x)\rightarrow 0,
\end{align}
where $Z^{\epsilon}=(Z^{\epsilon}_1, \cdots, Z^{\epsilon}_n)^\top$ is a multivariate normally distributed random variable with mean zero and covariance matrix $\Sigma^{\epsilon}$.
To this end,
let \begin{align*}
	\tilde T_{\tau}^{\epsilon0}=\max_{1\le i\le n} \frac{(\hat{p}_{i}^{\epsilon,bet}-\hat{p}^{\epsilon,in}_{i})^2}{(\frac{1}{n-n_{g_i}}+\frac{1}{n_{g_i}-1})(p^{\epsilon}_{22}-p^{\epsilon}_{12})},
\end{align*}
~where~ $\hat{p}_{i}^{\epsilon,bet}=\frac{1}{n-n_{g_i}}\sum_{j\not\in\mathcal{C}_{g_i}}I(||\epsilon_i-\epsilon_j||< \tau)$ and
$\hat{p}_{i}^{\epsilon, in}=\frac{1}{n_{g_i}-1}\sum_{j\in\mathcal{C}_{g_i}}I(||\epsilon_i-\epsilon_j||< \tau)$.
By Theorem 1 and Conditions (C1)--(C3), we  have
$P\big(\tilde T^{\epsilon0}_{\tau}<x\big)-P(\max_{1\le i\leq n} (Z_i^{\epsilon})^2<x)\rightarrow 0.$
Hence, to prove (\ref{A.20}), it suffices to show that $\tilde T_\tau^{\epsilon0}-\tilde{T}_\tau^{\epsilon}=o_p(1)$.

After algebraic calculation, we
obtain that
\begin{align}
	|\tilde T_{\tau}^{\epsilon0}-\tilde T^{\epsilon}_{\tau}|=&\left|\max_{1\le i\le n} \frac{(\hat{p}_{i}^{\epsilon,bet}-\hat{p}^{\epsilon, in}_{i})^2}{(\frac{1}{n-n_{g_i}}+\frac{1}{n_{g_i}-1})(p^{\epsilon}_{22}-p^{\epsilon}_{12})}-\max_{1\le i\le n} \frac{(\hat{p}_{i}^{\hat\epsilon,bet}-\hat{p}^{\hat\epsilon,in}_{i})^2}{(\frac{1}{n-n_{g_i}}+\frac{1}{n_{g_i}-1})(p^{\epsilon}_{22}-p^{\epsilon}_{12})}\right|\n\\
	\le &2\left|\max_{1\le i\le n} \frac{\hat{p}_{i}^{\epsilon,bet}-\hat{p}_{i}^{\hat\epsilon,bet}+\hat{p}^{\hat\epsilon,in}_{i}-\hat{p}^{\epsilon, in}_{i}}{\sqrt{(\frac{1}{n-n_{g_i}}+\frac{1}{n_{g_i}-1})(p^{\epsilon}_{22}-p^{\epsilon}_{12})}}\right|\max_{1\le i\le n} \frac{|\hat{p}_{i}^{\epsilon,bet}-\hat{p}^{\epsilon, in}_{i}|}{\sqrt{(\frac{1}{n-n_{g_i}}+\frac{1}{n_{g_i}-1})(p^{\epsilon}_{22}-p^{\epsilon}_{12})}}\n\\
	&+\max_{1\le i\le n} \frac{(\hat{p}_{i}^{\epsilon,bet}-\hat{p}_{i}^{\hat\epsilon,bet}+\hat{p}^{\hat\epsilon, in}_i-\hat{p}^{\epsilon, in}_{i})^2}{(\frac{1}{n-n_{g_i}}+\frac{1}{n_{g_i}-1})(p_{22}^{\epsilon}-p^{\epsilon}_{12})}. \label{A.21}
\end{align}
We next show that
\begin{align}
	\max_{1\le i\le n} \frac{|\hat{p}_{i}^{\epsilon,bet}-\hat{p}_{i}^{\hat\epsilon,bet}+\hat{p}^{\hat\epsilon,in}_{i}-\hat{p}^{\epsilon, in}_{i}|}{\sqrt{(\frac{1}{n-n_{g_i}}+\frac{1}{n_{g_i}-1})(p^{\epsilon}_{22}-p^{\epsilon}_{12})}}=O_p\Big(\sqrt{\frac{p\log n}{n}}\Big). \label{A.22}
\end{align}
It can be verified that
\begin{align*}
	&\max_{1\le i\le n} \frac{|\hat{p}_{i}^{\epsilon,bet}-\hat{p}_{i}^{\hat\epsilon,bet}+\hat{p}^{\hat\epsilon,in}_{i}-\hat{p}^{\epsilon, in}_{i}|}{\sqrt{(\frac{1}{n-n_{g_i}}+\frac{1}{n_{g_i}-1})(p^{\epsilon}_{22}-p^{\epsilon}_{12})}}\\
	&\le \max_{1\le i\le n} \frac{|\hat{p}_{i}^{\epsilon,bet}-\hat{p}_{i}^{\hat\epsilon,bet}|}{\sqrt{(\frac{1}{n-n_{g_i}}+\frac{1}{n_{g_i}-1})(p^{\epsilon}_{22}-p^{\epsilon}_{12})}}+\max_{1\le i\le n} \frac{|\hat{p}_{i}^{\epsilon, in}-\hat{p}_{i}^{\hat\epsilon,in}|}{\sqrt{(\frac{1}{n-n_{g_i}}+\frac{1}{n_{g_i}-1})(p^{\epsilon}_{22}-p^{\epsilon}_{12})}}.
\end{align*}
We are able to demonstrate that  the last two terms in the above equation are of order $O_p\big(\sqrt{\frac{p\log n}{n}}\big)$. Since their proofs are similar, we only present the proof of the second component.

It is worth noting that, with probability approaching 1, we obtain
\begin{align*}
	\hat{p}_{i}^{\epsilon, in}-\hat{p}_{i}^{\hat\epsilon,in}=&\frac{1}{n_{g_i}-1}\sum_{j\in\mathcal{C}_{g_i}}\big\{I(||\hat\epsilon_i-\hat\epsilon_j||< \tau)-I(||\epsilon_i-\epsilon_j||< \tau)\big\}\\
	&\le \frac{1}{n_{g_i}-1}\sum_{j\in\mathcal{C}_{g_i}}\Big\{I(||\epsilon_i-\epsilon_j||< \tau+||\hat\epsilon_i-\epsilon_i||+||\hat\epsilon_j-\epsilon_j||)\\
	&-I(||\epsilon_i-\epsilon_j||< \tau)\Big\}\\
	&\le C_1^* \frac{1}{n_{g_i}-1}\sum_{j\in\mathcal{C}_{g_i}}\Big\{||\hat\epsilon_i-\epsilon_i||+||\hat\epsilon_j-\epsilon_j||\Big\}\\ &=C_1^*||\hat\epsilon_i-\epsilon_i||+\frac{C_1^*}{n_{g_i}-1}\sum_{j\in\mathcal{C}_{g_i}}||\hat\epsilon_j-\epsilon_j||,
\end{align*}
where $C_1^*$ is a finite positive constant.
Accordingly,
\begin{align} {\label{A.23}}
	\max_{1\le i\le n}(\hat{p}_{i}^{\epsilon, in}-\hat{p}_{i}^{\hat\epsilon,in})\le C_1^*\max_{1\le i\le n}||\hat\epsilon_i-\epsilon_i||.
\end{align}
Analogously,  we have
\begin{align*}
	\hat{p}_{i}^{\hat\epsilon, in}-\hat{p}_{i}^{\epsilon,in}=&\frac{1}{n_{g_i}-1}\sum_{j\in\mathcal{C}_{g_i}}\left\{I(||\hat\epsilon_i-\hat\epsilon_j||< \tau)-I(||\epsilon_i-\epsilon_j||< \tau)\right\}\\
	&\ge \frac{1}{n_{g_i}-1}\sum_{j\in\mathcal{C}_{g_i}}\Big\{I(||\epsilon_i-\epsilon_j||< \tau-||\hat\epsilon_i-\epsilon_i||-||\hat\epsilon_j-\epsilon_j||)\\
	&-I(||\epsilon_i-\epsilon_j||< \tau)\Big\}\\
	&\ge -C_2^* \frac{1}{n_{g_i}-1}\sum_{j\in\mathcal{C}_{g_i}}\Big\{||\hat\epsilon_i-\epsilon_i||+||\hat\epsilon_j-\epsilon_j||\Big\}\\
	&=-C_2^*||\hat\epsilon_i-\epsilon_i||-\frac{C_2^*}{n_{g_i}-1}\sum_{j\in\mathcal{C}_{g_i}}||\hat\epsilon_j-\epsilon_j||,
\end{align*}
with probability approaching 1, where $C_2^*$ is a finite positive constant.
Thus,
\begin{align} {\label{A.24}}
	\max_{1\le i\le n}(\hat{p}_{i}^{\epsilon, in}-\hat{p}_{i}^{\hat\epsilon, in})\ge -C_2^*\max_{1\le i\le n}||\hat\epsilon_i-\epsilon_i||.
\end{align}
By (\ref{A.23}) and (\ref{A.24}), we have
\begin{align*}
	\max_{1\le i\le n}|\hat{p}_{i}^{\hat\epsilon, in}-\hat{p}_{i}^{\epsilon, in}|\le \max\{C_1^*, C_2^*\} \max_{1\le i\le n}||\hat\epsilon_i-\epsilon_i||.
\end{align*}
Using the ratio consistency of the least squares estimator,
we  further have $\max_{1\le i\le n}||\hat\epsilon_i-\epsilon_i||=O_p\big(\sqrt{\frac{p\log n}{n}}\big)$ by Conditions (C2) and (C3).
Accordingly,
\[
\max_{1\le i\le n} \frac{|\hat{p}_{i}^{\hat\epsilon, in}-\hat{p}_{i}^{\epsilon, in}|}{\sqrt{(\frac{1}{n-n_{g_i}}+\frac{1}{n_{g_i}-1})(p^{\epsilon}_{22}-p^{\epsilon}_{12})}}=O_p\Big(\frac{\sqrt{p\log n}}{n^{1/2}}\Big),\]
which verifies (\ref{A.22}). This, together with (\ref{A.21}),  implies that $|\tilde T^{\epsilon0}_{\tau}-\tilde T_{\tau}^{\epsilon}|=O_p\big(\frac{\sqrt{p\log n}}{n^{1/2}}\big)=o_p(1)$,
which completes the proof.

\section*{Section S.3: Derivation of $\sigma_{ij}$}

We consider the following two scenarios in deriving $\sigma_{ij}=E(Q_{i,\tau}Q_{j,\tau})$.
The first scenario is that $i$ and $j$ belong to the same sample for $i,j=1,\cdots,n$. Under this scenario, we obtain that
\begin{align*}
	E(Q_{i,\tau}Q_{j,\tau})=&\frac{E\big\{(\hat{p}_i^{bet}-\hat{p}^{in}_i)(\hat{p}_j^{bet}-\hat{p}^{in}_j)\big\}}{(\frac{1}{n-n_{g_{i}}}+\frac{1}{n_{g_{i}}-1})( p_{22}- p_{21})}\\
	=&\frac{1}{(\frac{1}{n-n_{g_{i}}}+\frac{1}{n_{g_{i}}-1})( p_{22}- p_{21})}\\
	&\times E\Bigg[\Big\{\frac{1}{n-n_{g_i}}\sum_{l\not\in\mathcal{C}_{g_i}}I(||X_i-X_l||\leq\tau)-\frac{1}{n_{g_i}-1}\sum_{l\in\mathcal{C}_{g_i}, l\not=i}I(||X_i-X_l||\leq\tau)\Big\}\\
	&\times \Big\{\frac{1}{n-n_{g_j}}\sum_{l\not\in\mathcal{C}_{g_j}}I(||X_j-X_l||\leq\tau)-\frac{1}{n_{g_j}-1}\sum_{l\in\mathcal{C}_{g_j}, l\not=j}I(||X_j-X_l||\leq\tau)\Big\}\Bigg]\\
	=&\frac{1}{(\frac{1}{n-n_{g_{i}}}+\frac{1}{n_{g_{i}}-1})( p_{22}- p_{21})}\frac{1}{(n-n_{g_i})^2}\sum_{l\not\in\mathcal{C}_{g_i}}\sum_{k\not\in\mathcal{C}_{g_i}}E(\delta_{il}\delta_{jk})\\
	&-\frac{1}{(\frac{1}{n-n_{g_{i}}}+\frac{1}{n_{g_{i}}-1})( p_{22}- p_{21})}\frac{2}{(n-n_{g_i})(n_{g_i}-1)}\sum_{l\not\in\mathcal{C}_{g_i}}\sum_{k\in\mathcal{C}_{g_j}, k\not=j}E(\delta_{il}\delta_{jk})\\
	&+\frac{1}{(\frac{1}{n-n_{g_{i}}}+\frac{1}{n_{g_{i}}-1})( p_{22}- p_{21})}\frac{1}{(n_{g_i}-1)^2}\sum_{l\in\mathcal{C}_{g_i}, l\not=i}\sum_{k\in\mathcal{C}_{g_j}, k\not=j}E(\delta_{il}\delta_{jk})\\
	=&\frac{1}{(\frac{1}{n-n_{g_{i}}}+\frac{1}{n_{g_{i}}-1})( p_{22}- p_{21})}\frac{1}{(n-n_{g_i})^2}\sum_{l\not\in\mathcal{C}_{g_i}}E(\delta_{il}\delta_{jl})\\
	&-\frac{1}{(\frac{1}{n-n_{g_{i}}}+\frac{1}{n_{g_{i}}-1})( p_{22}- p_{21})}\frac{2}{(n-n_{g_i})(n_{g_i}-1)}\sum_{l\not\in\mathcal{C}_{g_i}}E(\delta_{il}\delta_{ji})\\
	&+\frac{1}{(\frac{1}{n-n_{g_{i}}}+\frac{1}{n_{g_{i}}-1})( p_{22}- p_{21})}\frac{1}{(n_{g_i}-1)^2}E(\delta_{ij}^2)\\
	&+\frac{1}{(\frac{1}{n-n_{g_{i}}}+\frac{1}{n_{g_{i}}-1})( p_{22}- p_{21})}\frac{1}{(n_{g_i}-1)^2}\sum_{l\in\mathcal{C}_{g_i}, l\not=i,j}E(\delta_{il}\delta_{jl})\\
	&+\frac{1}{(\frac{1}{n-n_{g_{i}}}+\frac{1}{n_{g_{i}}-1})( p_{22}- p_{21})}\frac{2}{(n_{g_i}-1)^2}\sum_{l\in\mathcal{C}_{g_i}, l\not=i,j}E(\delta_{ij}\delta_{jl})\\
	=&\frac{\{\frac{1}{n-n_{g_{i}}}+\frac{1}{n_{g_{i}-1}}\}p_{12}-\frac{3p_{12}-p_{22}}{(n_{g_{i}}-1)^2}}{(\frac{1}{n-n_{g_{i}}}+\frac{1}{n_{g_{i}}-1})(p_{22}-p_{21})}.
\end{align*}

The second scenario is that $i$ and $j$ do not belong to the same sample for $i,j=1,\cdots,n$. Under this scenario, we have that
\begin{align*}
	E(Q_{i,\tau}Q_{j,\tau})=&\frac{1}{(\frac{1}{n-n_{g_{i}}}+\frac{1}{n_{g_{i}}-1})( p_{22}- p_{21})}\frac{1}{(n-n_{g_i})^2}\sum_{l\not\in\mathcal{C}_{g_i}}\sum_{k\not\in\mathcal{C}_{g_j}}E(\delta_{il}\delta_{jk})\\
	&-\frac{1}{(\frac{1}{n-n_{g_{i}}}+\frac{1}{n_{g_{i}}-1})( p_{22}- p_{21})}\frac{2}{(n-n_{g_i})(n_{g_i}-1)}\sum_{l\not\in\mathcal{C}_{g_i}}\sum_{k\in\mathcal{C}_{g_j}, k\not=j}E(\delta_{il}\delta_{jk})\\
	&+\frac{1}{(\frac{1}{n-n_{g_{i}}}+\frac{1}{n_{g_{i}}-1})( p_{22}- p_{21})}\frac{1}{(n_{g_i}-1)^2}\sum_{l\in\mathcal{C}_{g_i}, l\not=i}\sum_{k\in\mathcal{C}_{g_j}, k\not=j}E(\delta_{il}\delta_{jk})\\
	=&\frac{1}{(\frac{1}{n-n_{g_{i}}}+\frac{1}{n_{g_{i}}-1})( p_{22}- p_{21})}\frac{1}{(n-n_{g_i})^2}\sum_{l\not\in\mathcal{C}_{g_i}\cup \mathcal{C}_{g_j}}E(\delta_{il}\delta_{jl})\\
	&+\frac{1}{(\frac{1}{n-n_{g_{i}}}+\frac{1}{n_{g_{i}}-1})( p_{22}- p_{21})}\frac{2}{(n-n_{g_i})^2}\sum_{k\not\in\mathcal{C}_{g_j},k\not=i}E(\delta_{ij}\delta_{jk})\\
	&+\frac{1}{(\frac{1}{n-n_{g_{i}}}+\frac{1}{n_{g_{i}}-1})( p_{22}- p_{21})}\frac{1}{(n-n_{g_i})^2}E(\delta_{ij}^2)\\
	&-\frac{1}{(\frac{1}{n-n_{g_{i}}}+\frac{1}{n_{g_{i}}-1})( p_{22}- p_{21})}\frac{2}{(n-n_{g_i})(n_{g_i}-1)}\sum_{k\in\mathcal{C}_{g_j}, k\not=j}E(\delta_{ik}\delta_{jk})\\
	&-\frac{1}{(\frac{1}{n-n_{g_{i}}}+\frac{1}{n_{g_{i}}-1})( p_{22}- p_{21})}\frac{2}{(n-n_{g_i})(n_{g_i}-1)}\sum_{k\in\mathcal{C}_{g_j}, k\not=j}E(\delta_{ij}\delta_{jk})\\
	=&\frac{(n_{g_{i}}-1)^{1/2}(n_{g_{j}}-1)^{1/2}\{(p_{22}-(n+2)p_{12})\}}{(n-n_{g_{i}})^{1/2}(n-n_{g_{j}})^{1/2}(n-1)(p_{22}-p_{12})}.
\end{align*}

\section*{Section S.4: Five Competing Methods}

In this section, we introduce the five competing methods considered in simulation studies.
(1) The edge count test (abbreviated as EC) proposed by Friedman and Rafsky (1979). This test generalizes the Wald-Wolfowitz's runs test to multivariate
data, and it is implemented via two steps.
First, construct an undirected similarity graph in terms of usual "closeness" between  observations by using $L_2$ or $L_1$ distance,
such as the minimal spanning tree (MST) of the pooled sample observations.
Second, count the number of edges that
connect observations from different samples in the MST, and then construct the test.
(2) The generalized edge count test (abbreviated as GEC) proposed by
Chen and Friedman (2017). This test involves two steps. The first step is the same as that of the EC test.
The second step is proposing a Hotelling's $T^2$ type statistic to test whether the number of edges connecting two observations
from the same sample are equal to its theoretical expectation under the null hypothesis. (3)
The weighted edge count test (abbreviated as WEC) proposed by Chen et al. (2018).
This test is constructed by assigning within-sample edges with different weights.
Specifically, the assigned weights are based on which sample they are coming from rather than treating them equally in the edge-count test.
(4) The Graph-based LP-nonparametric test (abbreviated as GLP) proposed by Mukhopadhyay and Wang (2020).
This test is based on the spectral graph partitioning method applied to $K$-samples.
It rejects the null hypothesis of equal distribution
when the dependence between true group labels and estimated group labels are
statistically significant. (5) The multi-sample Mahalanobis crossmatch test (MMCM) proposed by Mukherjee et al. (2022). This test
is a graphical method based on the minimum non-bipartite matching. Accordingly, it can assess the closeness between
the distributions by utilizing the number of edges connecting data points from different classes.

\section*{Section S.5: Additional Simulation Results}

	This section includes three parts. Part I presents some simulation studies with different $\tau$s,
	Part II presents some simulation results for small $p$, and Part III reports simulation results in regression settings when $K=2$ and 6.

\noindent{\textbf{Part I: Simulation Results for Different $\tau$s.}}

In this part, we conduct simulation studies with the three different $\tau$s (i.e.,  the empirical 25\%, 50\%, and 75\% quantiles of the $\|X_i-X_j\|$s, for $i,j=1,\cdots, n$, where $i\not=j$). The simulation settings are the same as those in Section 4 of the main paper.
Tables S.1 and S.2 present the empirical sizes and powers, respectively, for $K=2$, and Tables S.3 and S.4 report the
results for $K=6$. Based on the results in these four tables, we have the following findings.
First, both MOD and CA-MOD tests control the sizes well under the three different choices of $\tau$s.
Second, the powers of these two tests are quantitatively comparable, while the $\tau$ to be the median of $\|X_i-X_j\|$s for $i,j=1,\cdots, n$
performs the best in most cases, which support our discussion about the choice of $\tau$ below Theorem 2 and we recommend it in the main paper.

\noindent{\textbf{Part II: Simulation Results with Small $p$s.}}

We present a simulation study of tests MOD and CA-MOD  with relatively small $p$s (i.e., $p=2$, 5 and 10), and simulation
settings are the same
as those in Section 4 of the main paper.
Table S.5 summarizes the empirical sizes, and both MOD and CA-MOD tests control size well.
Hence, these two tests are applicable for both high dimensional and fixed dimensional data.

\noindent{\textbf{Part III: Simulation Results in Regression Settings.}}

This part  conducts simulation studies of tests MOD and CA-MOD in regression settings when $K=2$ and 6.
The simulation settings are the same
as those in Section 4 of the main paper.   Table S.6 reports of the empirical powers of the  seven tests when $K=2$, and S.7 presents the empirical sizes and powers of the four tests when $K=6$. The illustrations of simulation findings can be found in Section 4 of the main paper.

\begin{table}[htbp!]
	\noindent
	{Table S.1: The empirical sizes of the MOD and CA-MOD tests with three different choices of $\tau$s (i.e., the  25\%, 50\% and 75\% quantiles of $\|X_i-X_j\|$s for $i,j=1,\cdots, n$)
		in
		the three different cases (i.e., mean shift, covariance shift and distribution shift alternatives) with $K=2$.}
	\begin{center}
		\vspace{0.28 cm}
		\begin{tabular}{cccc|cccc}
			\hline\hline
			Setting &$n$ & Quantile & Methods & $p=50$ & $p=100$ & $p=200$  &  $p=500$ \\
			\hline\hline
			Size-all cases  & 150  & 0.5  & MOD         &    0.047 &    0.056 &    0.049 &    0.051  \\
			&      &      & CA-MOD      &    0.044 &    0.040 &    0.044 &    0.046 \\
			&      & 0.25 & MOD         &    0.040 &    0.050 &    0.041 &    0.052 \\
			&      &      & CA-MOD      &    0.048 &    0.034 &    0.035 &    0.049 \\
			&      & 0.75 & MOD         &    0.036 &    0.037 &    0.034 &    0.042 \\
			&      &      & CA-MOD      &    0.037 &    0.048 &    0.047 &    0.047  \\
			\hline\hline
			& 300  & 0.5  & MOD         &    0.047 &    0.037 &    0.059 &    0.045 \\
			&      &      & CA-MOD      &    0.039 &    0.046 &    0.046 &    0.050 \\
			&      & 0.25 & MOD         &    0.045 &    0.045 &    0.041 &    0.037  \\
			&      &      & CA-MOD      &    0.038 &    0.046 &    0.046 &    0.051 \\
			&      & 0.75 & MOD         &    0.036 &    0.033 &    0.036 &    0.039 \\
			&      &      & CA-MOD      &    0.050 &    0.040 &    0.055 &    0.046  \\
			\hline\hline
		\end{tabular}
	\end{center}
\end{table}

\begin{table}[htbp!]
	\noindent{Table S.2:
		The empirical powers  of the MOD and CA-MOD tests with with three different choices of $\tau$s (i.e., the  25\%, 50\% and 75\% quantiles of $\|X_i-X_j\|$s for $i,j=1,\cdots, n$) in
		the three different cases (i.e., mean shift, covariance shift and distribution shift alternatives) with $K=2$.}
	\begin{center}
       \renewcommand{\arraystretch}{0.5} 
		\vspace{0.28 cm}
		\begin{tabular}{cccc|cccc}
			\hline\hline
			Setting &$n$ & Quantile & Methods & $p=50$ & $p=100$ & $p=200$  &  $p=500$ \\
			\hline\hline
			Case 1          & 150  & 0.5  & MOD         &    0.959 &    0.679 &    0.376 &    0.180 \\
			&      &      & CA-MOD      &    0.949 &    0.656 &    0.362 &    0.193 \\
			&      & 0.25 & MOD         &    0.959 &    0.675 &    0.324 &    0.122 \\
			&      &      & CA-MOD      &    0.901 &    0.594 &    0.308 &    0.138  \\
			&      & 0.75 & MOD         &    0.929 &    0.572 &    0.265 &    0.103  \\
			&      &      & CA-MOD      &    0.941 &    0.644 &    0.355 &    0.136 \\
			\hline\hline
			& 300  & 0.5  & MOD         &    1.000 &    0.990 &    0.793 &    0.322 \\
			&      &      & CA-MOD      &    1.000 &    0.991 &    0.787 &    0.334 \\
			&      & 0.25 & MOD         &    1.000 &    0.992 &    0.781 &    0.277 \\
			&      &      & CA-MOD      &    1.000 &    0.979 &    0.792 &    0.286 \\
			&      & 0.75 & MOD         &    1.000 &    0.984 &    0.702 &    0.231 \\
			&      &      & CA-MOD      &    1.000 &    0.983 &    0.731 &    0.281 \\
			\hline\hline
			Case 2          & 150  & 0.5  & MOD         &    0.379 &    0.392 &    0.425 &    0.427 \\
			&      &      & CA-MOD      &    0.644 &    0.663 &    0.683 &    0.694 \\
			&      & 0.25 & MOD         &    0.309 &    0.302 &    0.332 &    0.340   \\
			&      &      & CA-MOD      &    0.649 &    0.673 &    0.687 &    0.702 \\
			&      & 0.75 & MOD         &    0.299 &    0.316 &    0.340 &    0.346 \\
			&      &      & CA-MOD      &    0.539 &    0.557 &    0.566 &    0.585  \\
			\hline\hline
			& 300  & 0.5  & MOD         &    0.844 &    0.851 &    0.866 &    0.874 \\
			&      &      & CA-MOD      &    0.976 &    0.984 &    0.980 &    0.989 \\
			&      & 0.25 & MOD         &    0.672 &    0.704 &    0.700 &    0.705  \\
			&      &      & CA-MOD      &    0.990 &    0.984 &    0.988 &    0.989 \\
			&      & 0.75 & MOD         &    0.816 &    0.836 &    0.846 &    0.885   \\
			&      &      & CA-MOD      &    0.948 &    0.951 &    0.956 &    0.949  \\
			\hline\hline
			Case 3         & 150  & 0.5   & MOD         &    0.065 &    0.178 &    0.548 &    0.985 \\
			&      &      & CA-MOD      &    0.203 &    0.492 &    0.937 &    1.000 \\
			&      & 0.25 & MOD         &    0.055 &    0.157 &    0.498 &    0.895  \\
			&      &      & CA-MOD      &    0.172 &    0.442 &    0.886 &    0.965 \\
			&      & 0.75 & MOD         &    0.072 &    0.184 &    0.532 &    1.000 \\
			&      &      & CA-MOD      &    0.192 &    0.483 &    0.963 &    1.000 \\
			\hline\hline
			& 300  & 0.5  & MOD         &    0.194 &    0.653 &    0.990 &    1.000 \\
			&      &      & CA-MOD      &    0.350 &    0.897 &    1.000 &    1.000 \\
			&      & 0.25 & MOD         &    0.158 &    0.591 &    0.944 &    1.000 \\
			&      &      & CA-MOD      &    0.321 &    0.762 &    1.000 &    1.000  \\
			&      & 0.75 & MOD         &    0.185 &    0.674 &    1.000 &    1.000 \\
			&      &      & CA-MOD      &    0.342 &    0.906 &    1.000 &    1.000  \\
			\hline\hline
		\end{tabular}
	\end{center}
\end{table}

\begin{table}[htbp!]
	\noindent{Table S.3: The empirical sizes of the MOD and CA-MOD tests with three different choices of $\tau$s (i.e., the 25\%, 50\% and 75\% quantiles of $\|X_i-X_j\|$s for $i,j=1,\cdots, n$) in
		the three different cases (i.e., mean shift, covariance shift and distribution shift alternatives) with $K=6$.}
	\begin{center}
		\vspace{0.28 cm}
		\begin{tabular}{cccc|cccc}
			\hline\hline
			Setting &$n$ & Quantile & Methods & $p=50$ & $p=100$ & $p=200$  &  $p=500$ \\
			\hline\hline
			Size-all cases  & 300  & 0.5  & MOD         &    0.042 &    0.036 &    0.037 &    0.039 \\
			&      &      & CA-MOD      &    0.056 &    0.041 &    0.034 &    0.047 \\
			&      & 0.25 & MOD         &    0.064 &    0.059 &    0.059 &    0.058 \\
			&      &      & CA-MOD      &    0.068 &    0.067 &    0.068 &    0.055  \\
			&      & 0.75 & MOD         &    0.064 &    0.059 &    0.056 &    0.058  \\
			&      &      & CA-MOD      &    0.066 &    0.066 &    0.067 &    0.057 \\
			\hline\hline
			& 600  & 0.5  & MOD         &    0.043 &    0.039 &    0.049 &    0.038 \\
			&      &      & CA-MOD      &    0.051 &    0.037 &    0.043 &    0.048 \\
			&      & 0.25 & MOD         &    0.052 &    0.049 &    0.053 &    0.051  \\
			&      &      & CA-MOD      &    0.066 &    0.066 &    0.063 &    0.065 \\
			&      & 0.75 & MOD          &    0.055 &    0.043 &    0.043 &    0.048  \\
			&      &      & CA-MOD      &    0.068 &    0.057 &    0.045 &    0.061 \\
			\hline\hline
		\end{tabular}
	\end{center}
\end{table}

\begin{table}[htbp!]
	\noindent{Table S.4:
		The empirical powers  of the MOD and CA-MOD tests with three different choices of $\tau$s (i.e., the 25\%, 50\% and 75\% quantiles of $\|X_i-X_j\|$s for $i,j=1,\cdots, n$) in
		the three different cases (i.e., mean shift, covariance shift and distribution shift alternatives) with $K=6$.}
	\begin{center}
		\vspace{0.28 cm}
           \renewcommand{\arraystretch}{0.5} 
		\begin{tabular}{cccc|cccc}
			\hline\hline
			Setting &$n$ & Quantile & Methods & $p=50$ & $p=100$ & $p=200$  &  $p=500$ \\
			\hline\hline
			Case 1          & 300  & 0.5  & MOD         &    0.948 &    0.608 &    0.368 &    0.209 \\
			&      &      & CA-MOD      &    0.940 &    0.601 &    0.369 &    0.217 \\
			&      & 0.25 & MOD         &    0.947 &    0.695 &    0.372 &    0.199  \\
			&      &      & CA-MOD      &    0.902 &    0.638 &    0.332 &    0.199 \\
			&      & 0.75 & MOD         &    0.822 &    0.567 &    0.214 &    0.162 \\
			&      &      & CA-MOD      &    0.817 &    0.540 &    0.234 &    0.182 \\
			\hline\hline
			& 600  & 0.5  & MOD         &    1.000 &    0.999 &    0.823 &    0.523\\
			&      &      & CA-MOD      &    1.000 &    0.997 &    0.855 &    0.538 \\
			&      & 0.25 & MOD         &    1.000 &    1.000 &    0.861 &    0.633  \\
			&      &      & CA-MOD      &    1.000 &    1.000 &    0.873 &    0.644\\
			&      & 0.75 & MOD         &    1.000 &    0.867 &    0.756 &    0.437 \\
			&      &      & CA-MOD      &    1.000 &    0.889 &    0.775 &    0.449 \\
			\hline\hline
			Case 2          & 300  & 0.5  & MOD         &    0.401 &    0.421 &    0.436 &    0.442 \\
			&      &      & CA-MOD      &    0.574 &    0.586 &    0.592 &    0.604 \\
			&      & 0.25 & MOD         &    0.343 &    0.369 &    0.379 &    0.372 \\
			&      &      & CA-MOD      &    0.435 &    0.453 &    0.465 &    0.450 \\
			&      & 0.75 & MOD         &    0.530 &    0.537 &    0.579 &    0.570 \\
			&      &      & CA-MOD      &    0.553 &    0.576 &    0.594 &    0.606 \\
			\hline\hline
			& 600  & 0.5  & MOD         &    0.801 &    0.817 &    0.839 &    0.842 \\
			&      &      & CA-MOD      &    0.963 &    0.965 &    0.968 &    0.972  \\
			&      & 0.25 & MOD         &    0.673 &    0.698 &    0.702 &    0.701 \\
			&      &      & CA-MOD      &    0.846 &    0.861 &    0.839 &    0.852  \\
			&      & 0.75 & MOD         &    0.900 &    0.908 &    0.908 &    0.933  \\
			&      &      & CA-MOD      &    0.951 &    0.962 &    0.962 &    0.968  \\
			\hline\hline
			Case 3          & 300  & 0.5  & MOD         &    0.166 &    0.410 &    0.706 &    0.955 \\
			&      &      & CA-MOD      &    0.255 &    0.656 &    0.971 &    1.000 \\
			&      & 0.25 & MOD         &    0.082 &    0.193 &    0.533 &    0.910 \\
			&      &      & CA-MOD      &    0.126 &    0.361 &    0.879 &    1.000 \\
			&      & 0.75 & MOD         &    0.390 &    0.631 &    0.767 &    0.913 \\
			&      &      & CA-MOD      &    0.298 &    0.649 &    0.968 &    1.000 \\
			\hline\hline
			& 600  & 0.5  & MOD         &    0.377 &    0.811 &    0.995 &    1.000 \\
			&      &      & CA-MOD      &    0.509 &    0.944 &    1.000 &    1.000\\
			&      & 0.25 & MOD         &    0.113 &    0.477 &    0.941 &    1.000 \\
			&      &      & CA-MOD      &    0.216 &    0.837 &    1.000 &    1.000 \\
			&      & 0.75 & MOD         &    0.678 &    0.945 &    0.993 &    1.000 \\
			&      &      & CA-MOD      &    0.559 &    0.956 &    1.000 &    1.000 \\
			\hline\hline
		\end{tabular}
	\end{center}
\end{table}

\begin{table}[htbp!]
	\noindent{Table S.5: The empirical sizes of tests MOD and CA-MOD  when $K=2$ and $p=2,5$ and 10.}
	\begin{center}
		\vspace{0.28 cm}
		\begin{tabular}{cc|ccc}
			\hline\hline
			$n$ &  Methods & $p=2$ & $p=5$ & $p=10$   \\
			\hline\hline
			150                 &MOD   &    0.036 &    0.036 &    0.041 \\
			&CA-MOD  &    0.055 &    0.040 &    0.050 \\
			\hline
			300                 &MOD   &    0.045 &    0.048 &    0.041 \\
			&CA-MOD  &    0.056 &    0.055 &    0.058 \\
			\hline\hline
		\end{tabular}
	\end{center}
\end{table}

\begin{table}[htbp!]
	\noindent{Table S.6: The empirical powers of the seven tests (EC, GEC, WEC, MMCM, GLP, MOD and CA-MOD) under
		the three different cases (i.e., mean shift, covariance shift and distribution shift alternatives) when $K=2$ in a regression setting.}
   \renewcommand{\arraystretch}{0.5} 
	\begin{center}
		\vspace{0.28 cm}
		\begin{tabular}{ccc|cccc}
			\hline\hline
			Setting &$n$ & Methods & $p=25$ & $p=50$ & $p=100$  &  $p=200$ \\
			\hline\hline
			Case 1  &150   &EC    &    0.889 &    0.665 &    0.325 &    0.109 \\
			&      &GEC   &    0.999 &    0.999 &    0.994 &    0.990 \\
			&      &WEC   &    0.999 &    0.994 &    0.969 &    0.955 \\
			&      &MMCM  &    0.911 &    0.862 &    0.736 &    0.706 \\
			&      &GLP   &    1.000 &    0.911 &    0.664 &    0.427 \\
			&      &MOD   &    0.998 &    0.947 &    0.693 &    0.371 \\
			&      &CA-MOD  &    1.000 &    0.949 &    0.675 &    0.331 \\
			\hline
			& 300  &EC    &    0.995 &    0.924 &    0.664 &    0.323 \\
			&      &GEC   &    1.000 &    1.000 &    1.000 &    0.999 \\
			&      &WEC   &    1.000 &    1.000 &    0.999 &    0.994 \\
			&      &MMCM  &    0.995 &    0.964 &    0.846 &    0.727 \\
			&      &GLP   &    1.000 &    1.000 &    0.962 &    0.721 \\
			&      &MOD   &    1.000 &    1.000 &    0.991 &    0.796 \\
			&      &CA-MOD  &    1.000 &    1.000 &    0.994 &    0.779 \\
			\hline\hline
			Case 2   & 150  &EC    &    0.048 &    0.044 &    0.051 &    0.048 \\
			&      &GEC   &    0.852 &    0.906 &    0.951 &    0.990 \\
			&      &WEC   &    0.467 &    0.578 &    0.709 &    0.798 \\
			&      &MMCM  &    0.215 &    0.307 &    0.428 &   0.595  \\
			&      &GLP   &    0.151 &    0.187 &    0.269 &   0.331 \\
			&      &MOD   &    0.463 &    0.525 &    0.603 &    0.653 \\
			&      &CA-MOD  &    0.734 &    0.737 &    0.796 &    0.801 \\
			\hline
			&300   &EC    &    0.051 &    0.062 &    0.058 &    0.048 \\
			&      &GEC   &    0.938 &    0.951 &    0.968 &    0.993 \\
			&      &WEC   &    0.664 &    0.782 &    0.847 &    0.939 \\
			&      &MMCM  &   0.286  &   0.301  &   0.369  &   0.502 \\
			&      &GLP   &    0.267 &    0.270 &   0.281  &   0.308 \\
			&      &MOD   &    0.866 &    0.886 &    0.893 &    0.923 \\
			&      &CA-MOD  &    0.983 &    0.985 &    0.987 &    0.987 \\
			\hline\hline
			Case 3   & 150  &EC    &    0.051 &    0.056 &    0.125 &    0.252 \\
			&      &GEC   &    0.371 &    0.411 &    0.665 &    0.956 \\
			&      &WEC   &    0.050 &    0.046 &    0.107 &    0.336 \\
			&      &MMCM  &    0.066 &    0.061 &    0.096 &    0.239 \\
			&      &GLP   &    0.025 &    0.069 &    0.097 &    0.152 \\
			&      &MOD   &    0.055 &    0.082 &    0.176 &    0.530 \\
			&      &CA-MOD  &    0.095 &    0.208 &    0.521 &    0.939 \\
			\hline
			& 300  &EC    &    0.085 &    0.197 &    0.504 &    0.847 \\
			&      &GEC   &    0.391 &    0.593 &    0.924 &    1.000 \\
			&      &WEC   &    0.065 &    0.065 &    0.146 &    0.347 \\
			&      &MMCM  &   0.055  &    0.086 &    0.131 &    0.278 \\
			&      &GLP   &  0.062   &    0.097 &    0.124 &    0.180 \\
			&      &MOD   &    0.078 &    0.228 &    0.649 &    0.990 \\
			&      &CA-MOD  &    0.130 &    0.350 &    0.895 &    1.000 \\
			\hline\hline
		\end{tabular}
	\end{center}
\end{table}

\begin{table}[htbp!]
	\noindent{Table S.7: The empirical sizes and powers of the four tests (MMCM, GLP, MOD and CA-MOD) under
		the three different cases (i.e., mean shift, covariance shift and distribution shift alternatives) when $K=6$ in a regression setting.}
	\begin{center}
		\vspace{0.28 cm}
   \renewcommand{\arraystretch}{0.5} 
		\begin{tabular}{ccc|cccc}
			\hline\hline
			Setting & $n$ &  Method & $p=25$ & $p=50$ & $p=100$  &  $p=200$ \\
			\hline
			Sizes    & 600     &MMCM     &    0.037 &    0.085  &    0.090 &    0.092 \\
			&         &GLP      &    0.051 &    0.062  &    0.075 &    0.082 \\
			&         &MOD      &    0.036 &    0.030 &    0.039 &    0.030 \\
			&         &CA-MOD   &    0.058 &    0.056 &    0.049 &    0.044 \\
			\hline
			& 900     &MMCM     &    0.055 &    0.045 &    0.038 &    0.064  \\
			&         &GLP      &    0.041 &    0.052 &    0.076 &    0.082  \\
			&         &MOD      &    0.029 &    0.038 &    0.030 &    0.035 \\
			&         &CA-MAD   &    0.061 &    0.060 &    0.048 &    0.046 \\
			
			\hline\hline
			Power-Case 1     & 600     &MMCM     &    0.972  &    0.908 &    0.626 &    0.508 \\
			&         &GLP      &    1.000  &    1.000 &    1.000 &    0.981 \\
			&         &MOD      &    1.000  &    0.917  &    0.496 &    0.371 \\
			&         &CA-MAD   &    1.000  &    0.949  &    0.590 &    0.393 \\
			\hline
			& 900     &MMCM     &    1.000 &   0.991  &   0.884  &    0.649 \\
			&         &GLP      &    1.000 &    1.000 &    1.000 &    1.000  \\
			&         &MOD      &    1.000 &    0.999 &    0.832 &    0.586 \\
			&         &CA-MOD   &    1.000 &    1.000 &    0.892 &    0.590 \\
			\hline\hline
			Power-Case 2     & 600     &MMCM     &    0.256 &    0.279 &    0.325 &    0.442 \\
			&         &GLP      &    0.551 &    0.742 &    0.857 &    0.893 \\
			&         &MOD      &    0.503 &    0.531 &    0.559 &    0.598 \\
			&         &CA-MOD   &    0.734 &    0.772 &    0.880 &    0.899 \\
			\hline
			& 900     &MMCM     &    0.295 &    0.361 &    0.497 &     0.678 \\
			&         &GLP      &    0.840 &    0.877 &    0.942 &     0.944 \\
			&         &MOD      &    0.762 &    0.794 &    0.801 &    0.815 \\
			&         &CA-MOD   &    0.959 &    0.954 &    0.962 &    0.961 \\
			\hline\hline
			Power-Case 3     & 600     &MMCM     &    0.070 &    0.065 &    0.082 &    0.101 \\
			&         &GLP      &    0.106 &    0.121 &    0.196 &    0.569 \\
			&         &MOD      &    0.131 &    0.363 &    0.771 &    0.993 \\
			&         &CA-MOD   &    0.169 &    0.475 &    0.952 &    1.000 \\
			\hline
			& 900     &MMCM     &    0.064 &    0.057 &    0.099 &    0.126 \\
			&         &GLP      &    0.085 &    0.118 &    0.407 &    0.832 \\
			&         &MOD      &    0.209 &    0.581 &    0.960 &    1.000 \\
			&         &CA-MOD   &    0.253 &    0.686 &    0.997 &    1.000 \\
			\hline\hline
		\end{tabular}
	\end{center}
\end{table}

\vskip 0.2in

\section*{References}
\begin{description}
	\newcommand{\enquote}[1]{``#1''}
	\expandafter\ifx\csname natexlab\endcsname\relax\def\natexlab#1{#1}\fi

	\bibitem[Andrews(1997)]{Andrews:1997}
	Andrews, D. (1997), \enquote{A conditional Kolmogorov test},
	\textit{Econometrica}, 65, 1097-1128.

	\bibitem[Bai and Saranadasa(1996)]{Bai:1996}
	Bai, Z. D. and Saranadasa, H. (1996), \enquote{Effect of high dimension: By an example of a two sample problem},
	\textit{Statistica Sinica}, 6, 311-329.

	\bibitem[Bickel(1969)]{Bickel:1969}
	Bickel, P.~J. (1969), \enquote{A distribution free version of the smirnov two sample test in the p-variate case},
	\textit{Annals of Mathematical Statistics}, 40, 1--23.

	\bibitem[Cai et al. (2014)]{clx2014}
	Cai, T., Liu, W. and Xia, Y. (2014) \enquote{Two-sample test of high dimensional means under dependence},
	\textit{Journal of the Royal Statistical Society Series B (Statistical Methodology)}, 76, 349-372.

	\bibitem[Carpenter et al. (2021)]{car2021}
	Carpenter, J., Lu, F. and Whitelaw, R.~F. (2021) \enquote{The real value of China's stock market},
	\textit{Journal of Financial Economics}, 139, 679--696.

	
	\bibitem[Chen et al.(2018)]{Chen:2018}
	Chen, H., Chen, X. and Su, Y. (2018), \enquote{A weighted edge-count two-sample test for multivariate and object data},
	\textit{Journal of the American Statistical Association}, 113, 1146-1155.

	\bibitem[Chen and Friedman(2017)]{Chen:2017}
	Chen, H. and Friedman, J.~H. (2017), \enquote{A new graph-based two-sample test for multivariate and object data},
	\textit{Journal of the American Statistical Association}, 112, 397-409.


	\bibitem[Chen and Zhang(2013)]{Chen:2013}
	Chen, H. and Zhang, N. (2013), \enquote{Graph-based tests for two-sample comparisons of categorical data},
	\textit{Statistica Sinica}, 23, 1479-1503.


	\bibitem[Chernozhukov et al. (2013)]{c2013}
	Chernozhukov, V., Denis, C. and Kengo, K. (2013).
	\enquote{Gaussian approximations and multiplier bootstrap for maxima of sums of high-dimensional random vectors},
	\textit{Annals of Statistics}, 41, 2786-2819.


	\bibitem[Chong et al. (2012)]{ch2012}
	Chong, T.~L., Lam, T.~H. and Yan, K.~M. (2012).
	\enquote{Is the Chinese stock market really inefficient?},
	\textit{China Economic Review}, 23, 122-137.

	\bibitem[Dicle and Levendis(2014)]{Dicle:2014}
	Dicle, M. and Levendis, J. (2014), \enquote{Day-of-the-week effect revisited: International evidence},
	\textit{Journal of Economics and Finance}, 38, 407-437.

	\bibitem[Eagle et al.(2009)]{Eagle:2009}
	Eagle, N., Pentland, A. and Lazer, D. (2009), \enquote{Inferring friendship network structure by using mobile phone data},
	\textit{Proceedings of the National Academy of Sciences of the USA}, 106, 15274-15278.
	
	\bibitem[{Fama and French (1993)}]{Fama:1993}
	Fama, E. and French, K. (1993).
	\enquote{Common risk factors in the returns on stocks and bonds,}
	\textit{Journal of Financial Economics} 33, 3-56.

	\bibitem[Feng et al.(2022)]{Feng:2022}
	Feng, L., Lan, W., Liu, B. and Ma, Y. (2022), \enquote{High-dimensional test for alpha in linear factor pricing models with sparse alternatives},
	\textit{Journal of Econometrics}, 229, 152-175.
	

	\bibitem[Friedman and Rafsky(1979)]{Friedman:1979}
	Friedman, J. and Rafsky, L. (1979), \enquote{Multivariate generalizations of the Wald-Wolfowitz and Smirnov two-sample tests},
	\textit{Annals of Statistics}, 7, 697-717.
	
	
	\bibitem[{Gibbons and Chakraborti(2011)}]{Chakraborti:2011}
	Gibbons, J. D. and Chakraborti, S. (2011), \textit{Nonparametric Statistical Inference},
	Springer.

	\bibitem[Gibbons and Hess(1981)]{Gibbons:1981}
	Gibbons, M. and Hess, P. (1981), \enquote{Day of the week effects and asset returns},
	\textit{Journal of Business}, 54, 579-596.


	\bibitem[Guo et al. (2014)]{Guo:2014}
	Guo, H., Zou, C., Wang, Z. and Chen. B. (2014), \enquote{Empirical likelihood for high-dimensional linear
		regression models}, \textit{Metrika}, 77, 921-945.


		\bibitem[Heller et al.(2016)]{Heller:2016}
		Heller, R., Heller, Y.,  Kaufman, S., Brill, B. and Gorfine, M. (2016), \enquote{Consistent distribution-free $K$-sample and independence tests for
			univariate random variables},
		\textit{Journal of Machine Learning Research}, 17, 1-54.

		\bibitem[Jiang et al.(2015)]{Jiang:2015}
		Jiang, B., Chao, Y. and Liu, J. (2015), \enquote{Non-parametric $K$-sample tests via dynamic slicing},
		\textit{Journal of the
			American Statistical Association}, 110, 642-653.

	\bibitem[Kohers and Pandey(2004)]{Kohers:2004}
	Kohers, G. and Pandey, V. (2004), \enquote{The disappearing day-of-the-week effect in the world's largest equity markets},
	\textit{Applied Economics Letters}, 11, 167-171.

	\bibitem[Lehmann(2004)]{Lehmann:2004}
	Lehmann, E. L. (2004), {\it Elements of Large-Sample Theory},
	Springer.

	\bibitem[Li and Chen(2012)]{Li:Chen:2012}
	Li, J. and Chen, S. (2012), \enquote{Two sample tests for high-dimensional covariance matrices},
	\textit{Annals of Statistics}, 40, 908-940.

	\bibitem[{Li et al.(2009)}]{Li:2009}
	Li, Q., Maasoumi, E. and Racine, J. S. (2009).
	\enquote{A nonparametric test for equality of distributions with mixed categorical and continuous data,}
	\textit{Journal of Econometrics}, 148, 186-200.

	\bibitem[Li et al.(2012)]{Li:2012}
	Li, R., Zhong, W. and Zhu, L. (2012), \enquote{Feature screening via distance correlation learning},
	\textit{Journal of the American Statistical Association}, 107, 1129-1139.

	\bibitem[Mukherjee et al.(2022)]{Mukherjee:2022}
	Mukherjee, S., Agarwal, D., Zhang, N.~R. and Bhattacharya, B. (2022), \enquote{Distribution-free multisample tests based on
		optimal matchings with applications to single Cell Genomics},
	\textit{Journal of the American Statistical Association}, 117, 627-638.

	\bibitem[Mukhopadhyay and Wang(2020)]{Mukhopadhyay:Wang:2020}
	Mukhopadhyay, S. and Wang, K. (2020), \enquote{A nonparametric approach to high-dimensional
		$k$-sample comparison problems},
	\textit{Biometrika}, 107, 555-572.
	
	\bibitem[Oja and Randles(2004)]{Oja:2004}
	Oja, H. and Randles, R. H. (2004), \enquote{Multivariate nonparametric tests},
	\textit{Statistical Science}, 19, 598-605.

	\bibitem[{Shao(2003)}]{Shao:2003}
	Shao, J. (2003), \textit{Mathematical Statistics, Second Edition}, Springer-Verlag New York.

	\bibitem[Sharpe(1964)]{Sharpe:1964}
	Sharpe, W. (1964), \enquote{Capital asset prices: A theory of market equilibrium under conditions of risk},
	\textit{Journal of Finance}, 19, 425-444.

		\bibitem[Wynne and Duncan(2022)]{Duncan:2022}
		Wynne, G. and Duncan, A. (2022), \enquote{A kernel two-sample test for functional data},
		\textit{Journal of Machine Learning Research}, 23, 1-51.

	\bibitem[Xue and Yao(2020)]{Xue:2020}
	Xue, K. and Yao, F. (2020), \enquote{Distribution and correlation-free two-sample test of high-dimensional means},
	\textit{Annals of Statistics}, 48, 1304-1328.

		\bibitem[Zhang et al.(2024)]{Zhang:2024}
		Zhang, J.~T., Guo, J. and Zhou, B. (2024), \enquote{Testing equality of several distributions in separable metric
			spaces: A maximum mean discrepancy based approach},
		\textit{Journal of Econometrics}, 239, 105286.

	\bibitem[Zhong et al.(2013)]{Zhong:2013}
	Zhong, P. S., Chen, S. X. and Xu, M. (2013), \enquote{Tests alternative to higher criticism for high-dimensional means under sparsity and column-wise dependence},
	\textit{Annals of Statistics}, 41, 2820-2851.
	
\end{description}

\end{document}